\pdfoutput=1

\documentclass[10pt,a4paper,oneside]{article}

\usepackage[top=0.85in,left=2.5cm,right=2.5cm,footskip=0.75in,]{geometry}

\usepackage{graphicx}

\usepackage[utf8]{inputenc}
\usepackage[T1]{fontenc}
\usepackage[english]{babel}
\usepackage[expansion=false]{microtype}
\makeatletter
  \g@addto@macro\@verbatim{\microtypesetup{act‌​ivate=false}}
\makeatother

\usepackage[aboveskip=1pt,labelfont=bf,labelsep=period,singlelinecheck=off,font=small,]{caption}
\usepackage[%
  listformat=subsimple,
  subrefformat=parens,
  font=footnotesize,
]{subcaption}
\makeatletter
\renewcommand*\subcaption@label{%
  \caption@withoptargs\subcaption@@label}
\makeatother
\usepackage[caption2]{ccaption}

\usepackage[numbers,sort&compress,merge,round]{natbib}

\usepackage{amssymb,amsfonts,amsmath}
\usepackage{nameref,hyperref}
\usepackage{cleveref}
\usepackage{xcolor}
\hypersetup{
  colorlinks,
  linkcolor={red!50!black},
  citecolor={blue!50!black},
  urlcolor={blue!80!black},
  pdfauthor={%
    Loizos Kounios, Jeff Clune, Kostas Kouvaris, G{\"u}nter P. Wagner, Mihaela Pavlicev, Daniel M. Weinreich, Richard A. Watson
  },
  pdftitle={Resolving the paradox of evolvability with learning theory: How evolution learns to improve evolvability on rugged fitness landscapes},
  pdfkeywords={evolvability,evo--devo,genotype--phenotype maps},
}

\usepackage{fixltx2e}

\newcommand{\figFilename}{} 
\newcommand{\symvect}[1]{\mathbf{#1}} 
\newcommand{\symgenot}{\ensuremath{\symvect{G}}} 
\newcommand{\symbeta}{\ensuremath{\symvect{B}}} 
\newcommand{\symphenot}{\ensuremath{\symvect{P}}} 
\newcommand{\symadultphenot}{\ensuremath{{\symvect{P}_a}}} 
\newcommand{\sympmadultphenot}{\ensuremath{\bar{\symvect{P}}_a}} 
\newcommand{\symdevotime}{\ensuremath{t_{\text{final}}}} 
\newcommand{\ldblbracket}{[\![}
\newcommand{\rdblbracket}{]\!]}

\begin{document}
\vspace*{0.15in}

\begin{flushleft}
{\Large
\textbf\newline{Resolving the paradox of evolvability with learning theory: How evolution learns to improve evolvability on rugged fitness landscapes}
}
\newline
\\
Loizos Kounios\textsuperscript{1,*},
Jeff Clune\textsuperscript{2},
Kostas Kouvaris\textsuperscript{1},
G{\"u}nter P.\ Wagner\textsuperscript{3},
Mihaela Pavlicev\textsuperscript{4},
Daniel M.\ Weinreich\textsuperscript{5},
Richard A.\ Watson\textsuperscript{1}
\\
\bigskip
\bf{1} ECS, University of Southampton, Southampton SO17 1BJ, United Kingdom
\\
\bf{2} University of Wyoming, Laramie, Wyoming, USA
\\
\bf{3} Ecology and Evolutionary Biology, Yale University, Connecticut 06477
\\
\bf{4} Cincinnati Children's Hospital Medical Center, Cincinnati, OH.
\\
\bf{5} Ecology and Evolutionary Biology/Center for Computational Molecular Biology, Brown University, Providence, Rhode Island
\\
\bigskip
* To whom correspondence should be addressed: \href{mailto:lk9g12@soton.ac.uk}{lk9g12@soton.ac.uk}.

\end{flushleft}

\section*{Abstract}
It has been hypothesized that one of the main reasons evolution has been able to produce such impressive adaptations is because it has improved its own ability to evolve -- ``the evolution of evolvability''.
Rupert Riedl, for example, an early pioneer of evolutionary developmental biology, suggested that the evolution of complex adaptations is facilitated by a developmental organization that is itself shaped by past selection to facilitate evolutionary innovation.
However, selection for characteristics that enable future innovation seems paradoxical: natural selection cannot favor structures for benefits they have not yet produced; and favoring characteristics for benefits that have already been produced does not constitute future innovation.
Here we resolve this paradox by exploiting a formal equivalence between the evolution of evolvability and learning systems.
We use the conditions that enable simple learning systems to generalize, i.e., to use past experience to improve performance on previously unseen, future test cases, to demonstrate conditions where natural selection can systematically favor developmental organizations that benefit future evolvability.
Using numerical simulations of evolution on highly epistatic fitness landscapes, we illustrate how the structure of evolved gene regulation networks can result in increased evolvability capable of avoiding local fitness peaks and discovering higher fitness phenotypes.
Our findings support Riedl's intuition: Developmental organizations that ``mimic'' the organization of constraints on phenotypes can be favored by short-term selection and also facilitate future innovation.
Importantly, the conditions that enable the evolution of such surprising evolvability follow from the same non-mysterious conditions that permit generalization in learning systems.

\section*{Introduction}
The ability of natural populations to exhibit adaptation depends on the production of suitable phenotypic variation that natural selection can act on.
Such variability in turn depends, amongst other things, on the properties of the genotype--phenotype map or the organization of developmental processes that constrain and bias the distribution of possible phenotypic variants.
Contemporary evolutionary developmental biology recognizes that developmental organization is both a product of natural selection and a factor that can significantly alter subsequent evolutionary outcomes~\cite{a:riedl:1977:01,b:riedl:1978:01,a:wagner:1996:01,a:toussaint:2007:01}.
It is therefore clear that natural selection can produce heritable changes that modify the ability to evolve -- a phenomenon called the ``evolution of evolvability''~\cite{a:wagner:1996:01,a:kirschner:1998:01,a:gerhart:2007:01}.
But can natural selection systematically improve its own ability to evolve?
It has been argued that natural selection increases evolvability over time and that this is important in explaining the amazing diversity and complexity of the natural world~\cite{a:riedl:1977:01,b:riedl:1978:01,ic:conrad:1998:01}.
Yet, how natural selection, which operates only on short-term fitness differences, can improve long-term evolvability remains one of the most important, open questions in evolutionary biology~\cite{a:conrad:1972:01,a:riedl:1977:01,b:riedl:1978:01,a:conrad:1979:01,a:conrad:1990:01,ic:altenberg:1995:01,a:wagner:1996:01,ic:conrad:1998:01,a:kirschner:1998:01,a:gerhart:2007:01,a:diaz-arenas:2013:01}.

Rupert Riedl, an early pioneer of evo--devo research, provided key concepts and ideas on the topic of evolvability.
He suggested that the evolution of complex adaptations is facilitated by developmental architectures that are organized by natural selection to ``mimic'' or ``imitate'' the functional constraints on phenotypes~\cite{a:riedl:1977:01,b:riedl:1978:01,a:wagner:1996:01,a:wagner:2004:01}.
However, his ideas have not been previously demonstrated in an explicit mechanistic model.
Such a model needs to explain exactly what form such imitation takes and, more challengingly, explain why imitating the constraints experienced in the past facilitates evolutionary innovation in the future.
We provide a resolution to this problem by drawing a formal analogy with well-established knowledge from another discipline; namely, learning theory.
A simple analogy between learning and evolution has been noted many times~\cite{b:skinner:1965:01,b:maynard-smith:1986:01}, but the link has recently been formalized and deepened extensively~\cite{ic:valiant:2007:01,a:watson:2014:01,a:chastain:2014:01,a:power:2015:01,a:kouvaris:2015:01,a:watson:2016:01,a:watson:2016:02}.
Here we argue that  Riedl's notion of an imitatory developmental organization is directly analogous to a learning system that internalizes a model of its environment, and the link with learning also provides the missing explanatory component and mechanistic principles to demonstrate how this facilitates future innovation.
Specifically, in learning systems the idea that past experience can be used to generalize to future, previously unseen, test cases is not at all paradoxical.
This enables us to update Riedl's notion of development that \emph{imitates} to a process of development that \emph{innovates} in a simple but predictable manner.

\subsection*{Generalization and evolvability}
Here, we develop the intuition that evolvability is to evolution as generalization is to learning~\cite{a:watson:2016:01,a:watson:2016:02}.
That is, learning without generalization is essentially just remembering what you have already seen; it is obvious that evolution by natural selection can do this in the sense that genotype frequencies are determined by past selection.
But evolvability suggests an ability to go beyond this; not merely to favor the phenotypes that have already been rewarded but to facilitate innovation -- an ability to produce new phenotypes that are fit even though they have not been previously subject to selection.
The evolution of a high mutation rate in a rugged fitness landscape is potentially a simple way to provide this -- implicitly exploiting the prediction that high degrees of epistasis in the past might be indicative of high degrees of epistasis in the future.
But a parameter that simply controls the amount of variability (e.g., mutation rate) has little or no ability to accumulate specific information from past selection and is, in any case, not easy to evolve~\cite{a:clune:2008:01}.
Learning systems show that predicting the future (in specific ways) from past experience is possible in simple mechanistic systems.
Learning systems do not really ``see the future'', of course; they are simply finding underlying structural regularities that are invariant over time~\cite{ic:valiant:2007:01}.
For this to be non-trivial, it is essential that the future is not simply the same as the past, but rather shares invariant structural properties -- properties that are, in some sense, beneath the surface.
In this paper, by transferring existing knowledge from learning systems, we explore the conditions where natural selection can achieve this same type of generalization and thus demonstrate  evolvability that facilitates innovation.
Unlike models based on the evolution of mutation rates that attempt to increase evolvability simply by enabling more (random) variability, our models increase evolvability by enabling ``smarter'' variability;  i.e., by evolving developmental organizations that create a distribution of variants with specific adaptive structure.

To learn structural regularities requires an ability to accumulate information from multiple past examples.
A learning system that models each feature independently cannot do this because each new example overwrites the features of the previous example (to the extent that they are different).
This is analogous to an evolutionary system where phenotypes are directly encoded by genotypes -- i.e., each phenotypic trait is controlled by a single gene, also known as ``one-to-one''.
Given that genetic variation is undirected, a one-to-one mapping produces undirected variability in phenotypic characteristics and is therefore incapable of ``remembering'' more than one thing (except in the degenerate sense of representing a simple average), or generalizing~\cite{a:kouvaris:2015:01,a:watson:2016:01}.
The evolutionary significance of this is that it is unable to use information from past selective environments to modify the future response to selection except by changing its \emph{current} phenotype -- there is no possibility of representing tacit information from past selection.
In contrast, a learning system that generalizes requires an ability to represent the underlying structural regularities of a problem domain, e.g., to observe which features ``go together'' in a set of examples, and this requires a more flexible input--output mapping.
Representing structural relationships between multiple features of observed solutions enables a learning system to exploit similarities between multiple past solutions (and possible future solutions) even though their individual features may be different or contradictory.

This is analogous to a G--P map that constrains and biases the combinations of phenotypic traits produced by genotypes, e.g., via a gene-regulatory network.
This is capable of representing information about structural regularities observed in past selected phenotypes and generalizing by facilitating phenotypic variation which reflects the regularities represented within the network~\cite{a:watson:2014:01,a:kouvaris:2015:01}.
Recent work shows that it is possible for a gene-regulation network to represent information about structural regularities observed in past selected phenotypes and also describes in detail how information is introduced into such networks by past selection in the same way that an associative learning system acquires information from experience~\cite{a:watson:2014:01}.
Specifically, this work shows that the effect of natural selection on mutations that change the strength of connections in a gene-regulation network is formally equivalent to learning mechanisms that alter the strength of synaptic connections in a neural network~\cite{a:watson:2014:01}.
The potential for neural networks to generalize via such simple learning mechanisms is well characterized.
Thus, a simple G--P mapping based on gene-regulatory interactions is capable of representing such correlations in the same way that a neural network is.
Here we show for the first time that this ability of natural selection to shape the organization of gene-networks can improve evolvability on rugged fitness landscapes, enabling future evolution to move in a modified space of phenotypes and thereby avoid local fitness peaks that prevent evolution with one-to-one mappings from finding higher fitness phenotypes.
We also show that the evolving gene network achieves this result by creating pleiotropic interactions that mimic the structure of the epistasis, inherent in the fitness landscape, acting on phenotypic traits, as Riedl predicted; but crucially, we also show that it is thus capable of generalization necessary to improve future innovation.
It is in this way that natural selection is able to recognize and exploit structural regularities in the selective environment that facilitate evolutionary innovation.

To explain this evolvability we need to explain the congruence between short-term fitness benefits (arising from organizations that mimic the structure of selective constraints on the phenotype) and long-term fitness benefits (i.e., an ability to innovate on rugged fitness landscapes).
This is neither guaranteed nor mysterious; it depends on the same conditions where generalization is possible in learning systems.
Specifically, that learnable regularity is present in the training data.
To demonstrate this, we utilize the observation that the phenotypes at local peaks in a multi-peaked epistatic fitness landscape are not arbitrarily different to one another (except in special cases).
It is, of course, possible to construct truly random landscapes where high-fitness phenotypes have no shared commonalities.
But this will not be the case whenever phenotypes have fitnesses that result from a sum of low-order epistatic fitness contributions (e.g., pairwise epistasis, but strictly, anything less than order-$n$ epistasis where $n$ is the number of traits).
In general, phenotypes at local fitness peaks share common structural regularities because they derive from the same set of underlying epistatic constraints~\cite{a:watson:2011:02,a:watson:2011:04}.
The presence of such regularities is of no use to natural selection with a one-to-one G--P map, since such a map is incapable of representing or exploiting them.
But we show that evolution on such a landscape, using phenotypes produced by an evolving gene-regulation network, can ``learn'' to exploit these regularities.
This learning--evolution analogy is more than the simple idea that selection ``rewards'' fit phenotypes by increasing their frequency in the population.
Rather this is a formal functional equivalence meaning that natural selection has the ability to induce an internal structural organization (in this case, via the evolution of gene-regulatory interactions) that mimics the structure of the epistasis acting on phenotypes.
This ``internalized model'' of the environment produces phenotypes that respect the epistatic interactions acting on those phenotypes and is thus enriched for fit phenotypes -- including novel phenotypes.
This causes future evolutionary trajectories to be biased toward parts of the fitness landscape with structural regularities that are ``familiar'' from past evolutionary experience.
This has the effect of causing future evolution to avoid low-fitness peaks (because they have less structural commonality with high-fitness phenotypes experienced in the past) and thus find high-fitness phenotypes faster and more reliably.
We demonstrate this by comparing evolutionary trajectories with more evolved and less evolved gene networks, from many different starting points in phenotype space, and measuring the long-term fitnesses they attain.
We are particularly interested in cases where evolutionary trajectories with a more evolved mapping systematically arrive at different local optima of higher fitness than evolutionary trajectories with a less evolved mapping.
Passing this test is important because it indicates that the more evolvable mapping enables the discovery of new high-fitness phenotypes that cannot be discovered with the less evolvable mapping (unless evolution moves against selective gradients or depends on multiple specific genetic mutations occurring simultaneously).
Accordingly, we compare multi-generation evolutionary trajectories with different G--P maps: (a) a direct encoding where the phenotype is the same as the genotype (i.e., a ``one-to-one'' or identity mapping); (b) evolved G--P maps at different stages of evolution.
Specifically, the direct encoding has no gene-regulatory interactions such that each gene controls exactly one phenotypic trait, whereas the evolved mappings introduce gene-regulatory interactions.
We assess the fitness levels attained and whether different fitness peaks (with phenotypes of higher fitness) are discovered in a multi-peaked fitness landscape.

Recent investigations into the evolution of evolvability have demonstrated that short-term natural selection can increase robustness~\cite{a:wagner:2005:01,a:wagner:2008:01,a:wagner:2012:01}, increase the rate of adaptation under directional selection~\cite{a:pavlicev:2011:01}, re-evolve previously evolved phenotypes more quickly~\cite{a:kashtan:2007:01}, and enable evolution to track changes in the environment more rapidly~\cite{a:lipson:2002:01,a:kashtan:2007:01,a:clune:2013:01}.
Some of this work is evidence of the imitatory features that Riedl described, e.g., modularity, or other structural features, that mimic the structure of the environment~\cite{a:parter:2008:01,a:clune:2013:01}.
Clune et al., in work on the evolution of neural networks, go further to demonstrate that conditions favoring modular networks not only mimic the modularity of the task environment but also facilitate more effective evolvability and superior evolutionary outcomes.
Here, investigating evolvability in a gene network (rather than a neural network), we show that there is sufficient learnable regularity in much more general fitness landscapes (without designed-in modularity).
More importantly, we explain the mechanism by formally linking the conditions for this result to generalization in learning theory.

\subsection*{Evolvability on rugged fitness landscapes}
To understand how evolutionary trajectories, starting from the same phenotype on the same fitness landscape, can be different from one another, we need to understand (a) that different G--P maps can produce different distributions of phenotypic variation~\cite{a:conrad:1972:01,a:riedl:1977:01,b:riedl:1978:01,a:conrad:1979:01,a:conrad:1990:01,ic:altenberg:1995:01,a:wagner:1996:01,ic:conrad:1998:01,a:kirschner:1998:01,a:gerhart:2007:01,a:toussaint:2007:01}; and (b) how this distribution interacts with selection to determine the direction of movement in phenotype space~\cite{a:arnold:2001:01,a:arnold:2008:01}.

Given that \emph{genetic} variation is undirected, a one-to-one mapping produces undirected variability in phenotypic characteristics also, and evolutionary trajectories are thus shaped by selective gradients only.
Mappings that produce non-uniform phenotypic distributions may cause differences in subsequent evolutionary trajectories~\cite{a:arnold:2001:01,a:arnold:2008:01}.
Under constant directional selection, if phenotypic variability in the direction of selection is decreased, this will decrease the rate of adaptation.
Conversely, if the phenotypic variation in the direction of selection is increased this will accelerate evolution~\cite{a:pavlicev:2011:01}.
Pavlicev et al.~\cite{a:pavlicev:2011:01} showed that the effect of short-term selection is to favor genetic lineages that increase phenotypic variation in the direction of selection precisely because they accelerate adaptation in this manner.
Moreover, biases in the distribution of phenotypic variants can not only increase or decrease the rate of adaptation but can also alter the direction of evolutionary trajectories through phenotype space~\cite{a:arnold:2001:01,a:arnold:2008:01}.
In much the same way that a keel on a sailing boat can make the direction of travel differ from the direction of the wind, a biased phenotypic distribution, by making evolutionary change in some dimensions easier than others, can make the path of evolutionary change differ from the current direction of selection and thus cause evolutionary trajectories to visit different regions of phenotype space.

In any single-peaked landscape, this may not alter long term evolutionary outcomes; assuming that a G--P map is not so constrained as to remove any of the necessary variability altogether, evolution will ultimately attain the same phenotypic peak and hence the same fitness in the long term, whether it be via a direct or circuitous route~\cite{a:arnold:2001:01,a:arnold:2008:01}.
In a multi-peaked fitness landscape, if a G--P map changes the regions of phenotype space that are explored, this causes natural selection to find different fitness peaks -- thus altering long-term evolutionary outcomes.
\Cref{fig:grn-cft100-cross-section-with-evolutionary-time-with-trajectories} demonstrates this using data from the experiments that follow.
The figure shows how evolution over a fixed set of possible mutations given one G--P map quickly exhausts the production of beneficial variation and becomes trapped at a local optimum in the adaptive landscape (\Cref{fig:grn-cft100-cross-section-with-evolutionary-time-with-trajectories:a}).
Nonetheless evolution over the same set of possible mutations with a different G--P map continues to produce adaptive variations that ultimately lead to superior regions of phenotype space (\Cref{fig:grn-cft100-cross-section-with-evolutionary-time-with-trajectories:b,fig:grn-cft100-cross-section-with-evolutionary-time-with-trajectories:c}).

\begin{figure}[p]
  \renewcommand{\figFilename}{grn-cft100-cross-section-with-evolutionary-time-with-trajectories}
  \centering
  \includegraphics[width=\textwidth,]{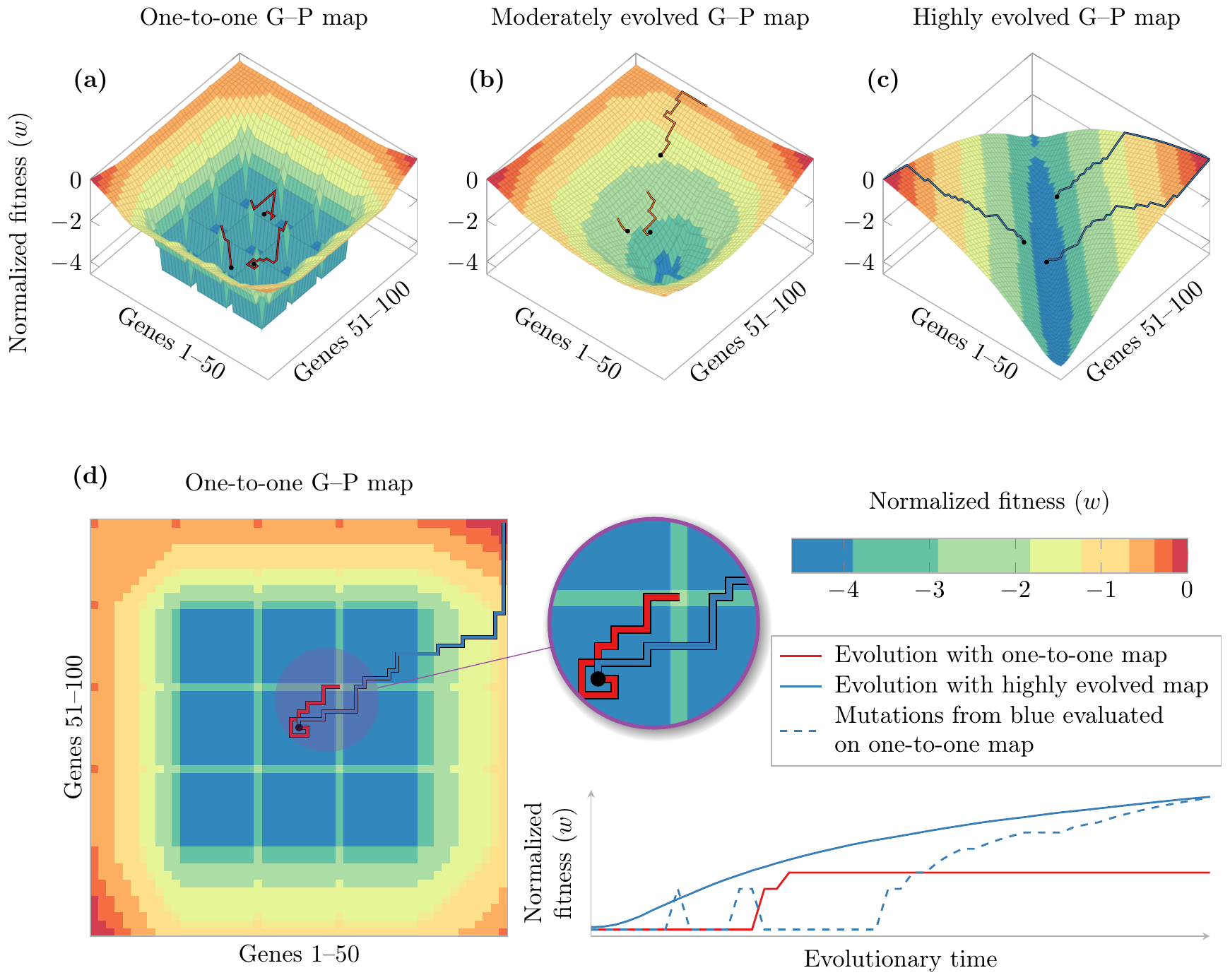}%
{%
  \ignorespaces\phantomsubcaption\label{fig:\figFilename:a}%
  \ignorespaces\phantomsubcaption\label{fig:\figFilename:b}%
  \ignorespaces\phantomsubcaption\label{fig:\figFilename:c}%
  \ignorespaces\phantomsubcaption\label{fig:\figFilename:d}%
}
  \caption{%
    Highly evolved G--P maps enable microevolution to find fitter phenotypes by changing the fitness landscape.
    The fitness landscape for a set of genetic mutations is modified by evolved changes in the genetic background defining the genotype--phenotype (G--P) map.
    Each mapping determines the phenotype (i.e., expression profile of $100$ gene products) produced by a set of $100$ genes.
    The initial mapping is a direct encoding (with no gene-regulatory interactions) such that each of the phenotypic traits is determined by the state of exactly one of the genes.
    Genetic mutations under this ``one-to-one'' or ``identity mapping'' have no pleiotropy or phenotypic epistasis, but the adaptive landscape exhibits widespread epistasis for fitness.
    The other mappings are the result of evolutionary changes (at other genetic sites) that introduce gene-regulatory interactions.
    In each case the surface shown depicts the same two-dimensional cross-section through the highly-epistatic high-dimensional fitness landscape (each dimension is a particular cross-section through a subspace of genotypes; see \emph{\nameref{subsec:cross-sections-through-genotype-space}}).
    This cross-section maintains the property that local optima on this surface (i.e., where all neighboring points are lower in fitness) are also true local optima in the original high-dimensional fitness landscape.
    The two globally optimal phenotypes in this landscape are located at opposite corners of this cross-section.
    Fitness reported is in logarithmic scale.
    \subref{fig:\figFilename:a}~With the one-to-one G--P map, the space of phenotypes corresponds directly to the underlying space of genotypes.
    Trajectories show simulation data of paths on the fitness surface taken by evolution using this direct encoding (start points drawn randomly in genotype space fall in low fitness regions of this surface with high probability).
    In this rugged fitness landscape these trajectories quickly become trapped when a local optimum is encountered.
    Evolution fails to find a global optimum in $1000$ attempts.
    Three example evolutionary trajectories are shown (black circles denote starting locations); to enhance visualization, trajectories which revisit previously seen phenotypes via neutral changes are not shown.
    \subref{fig:\figFilename:b}~The fitness surface with a moderately evolved G--P map.
    As the genetic background defining the parameters of the G--P map (i.e., evolving gene-regulatory interactions) evolve under natural selection, the phenotypes produced over the same space of $100$ genes change (in predictable ways, see text) and thus have different fitnesses.
    Some points in genotype space which were local optima in \subref{fig:\figFilename:a} are no longer local optima in \subref{fig:\figFilename:b}, and thus evolutionary trajectories from the same starting points as \subref{fig:\figFilename:a} sometimes discover higher fitness phenotypes.
    This is because the evolved regulatory interactions change the phenotypic neighborhood in a way which enables evolutionary trajectories to follow fitness gradients to different local optima.
    \subref{fig:\figFilename:c}~The fitness surface with a highly evolved G--P map.
    This mapping produces phenotypes that can always be improved by small mutations, and evolutionary trajectories thus reliably find the global optima when starting from any point ($1000$ random starting positions tested, all succeed).
    (Continued on next page.)
  }
  \label{fig:\figFilename}
\end{figure}

\begin{figure}[!t]
  \contcaption{%
    (Continued from previous page.)
    \subref{fig:grn-cft100-cross-section-with-evolutionary-time-with-trajectories:d}~Comparing the fitness effects of movements in the same genetic space given different mappings (enlarged area and line plots).
    The plot shows one trajectory from \subref{fig:grn-cft100-cross-section-with-evolutionary-time-with-trajectories:a} (red) and one trajectory from \subref{fig:grn-cft100-cross-section-with-evolutionary-time-with-trajectories:c} (blue), both overlaid on the fitness surface resulting from the G--P map in \subref{fig:grn-cft100-cross-section-with-evolutionary-time-with-trajectories:a}.
    The depicted evolutionary trajectory (red) gets stuck at a local optimum with low fitness.
    Trajectories through genotype space with the evolved mapping in \subref{fig:grn-cft100-cross-section-with-evolutionary-time-with-trajectories:c} (blue) include mutations that would have been deleterious with the direct encoding (see fitness decrease in dashed blue line plot).
    However, these genetic mutations are accepted by natural selection because they produce beneficial changes to phenotypes given the evolved mapping (see monotonically increasing fitness in blue line plot).
    Unlike the case with the one-to-one mapping, mutations that decrease the genetic distance to the fittest genotype always increase fitness with the evolved mapping.
    The evolved mappings achieve this via the evolution of regulatory interactions that bias the distribution of phenotypes produced by genetic mutations such that the expression of trait combinations that conflict with high-fitness phenotypes is suppressed and the expression of trait combinations that are compatible with high-fitness phenotypes is amplified.%
  }%
\end{figure}

To explain a systematic mechanism for the evolution of evolvability of this type requires that we understand the kind of knowledge that natural selection can ``learn'' from past experience (i.e., the kind of information that the organization of the G--P map can hold) and its correspondence with the knowledge that is needed to evolve high-fitness phenotypes.
This is what Riedl could not provide.
Here we show that when the G--P map is defined by a network of gene-regulatory connections, it is capable of learning the structure of pairwise epistatic interactions in the fitness landscape.
This biases the distribution of phenotypes produced by genetic mutations to suppress the expression of trait combinations that conflict with high-fitness phenotypes and amplify the expression of trait combinations that are compatible with high-fitness phenotypes.
These biases are a simple way of generalizing the information (combinations of phenotypic traits) found at multiple easy-to-find local peaks of average (for peaks) fitness, it is thereby able to predict the location of much rarer local peaks of exceptionally high fitness.

\subsection*{Experimental set-up}
We simulate the evolution of a GRN controlling a set of phenotypic traits, $\symphenot$, via a set of gene-expression potentials~\cite{a:watson:2014:01} (see \emph{\nameref{subsec:gene-regulation-network}}).
The genotype defines the embryonic state of each gene expression potential with a vector, $\symgenot$, and an interaction matrix, $\symbeta$, whose elements represent the magnitude and sign of the regulatory interactions between one gene and another in the GRN~\cite{a:watson:2014:01}.
In the absence of evolved regulatory interactions, $\symbeta$, the G--P map is one-to-one, and the sign and magnitude of elements in $\symphenot$ correspond directly to the elements of $\symgenot$.
Thus, in the absence of $\symbeta$, mutations to $\symgenot$ produce a uniform ball of variation in phenotype space on average.
In the presence of evolved regulatory interactions, $\symgenot$ is mapped into an adult phenotype using a non-linear, recurrent developmental process defined by $\symbeta$.

Developmental constraints, $\symbeta$, are assumed to evolve slowly.
Specifically, the correlations between traits, represented by the interaction matrix $\symbeta$, are assumed to evolve slowly relative to changes in the magnitudes of the traits themselves, represented by the elements of $\symgenot$ (e.g., the number of mutational sites affecting the expression level of a single gene is greater than the number of sites affecting its co-regulatory interaction with another specific gene~\cite{a:watson:2014:01}).
We refer to the timescale over which developmental constraints evolve as ``deep'' evolutionary time, and the more rapid timescale of evolution \emph{given} those developmental constraints as microevolutionary time.
We are interested in how the genetic constraints of the GRN (defined by $\symbeta$) change over deep time and how they modify the mapping between $\symgenot$ and $\symadultphenot$, the adult phenotype; and in particular, how the distribution of phenotypes produced under mutations to $\symgenot$ (for a given $\symbeta$) thus have the potential to modify microevolutionary trajectories through phenotype space.

It is not possible for evolution, or any other adaptive process, to infer the underlying structure of fitness a landscape from a single point in that landscape.
The prior work of Alon and colleagues thus elaborates the idea that varying goals -- changing repeatedly from one selective target to a different (but structurally similar) selective target -- enhances the evolution of evolvability to previously seen environments~\cite{a:kashtan:2007:01,a:parter:2008:01,a:crombach:2008:01,a:clune:2013:01} and in some cases (e.g., modularly-varying goals) produces generalization to previously-unseen environments~\cite{a:parter:2008:01,a:crombach:2008:01,a:clune:2013:01,a:watson:2014:01}.
In that work it was necessary to hand-design a family of single-peaked selective landscapes sharing common structural regularities.
Instead we utilize the observation that the local peaks present in a single multi-peaked landscape naturally share common structural regularities because they derive from the same set of underlying epistatic constraints~\cite{a:watson:2011:02,a:watson:2011:04}.
This does not depend on any explicit modularity or other contrivance in the problem structure.
This also has the advantage that the fitnesses of different local peaks are comparable because they are on the same landscape.
This is important because it enables us to assess not merely whether evolution evolves a given phenotype \emph{more quickly}, but also whether evolution can find a different phenotype that is \emph{fitter}.
This ability cannot be assessed when the landscape (at any one time) is single-peaked.

Thus, rather than changing between multiple single-peaked fitness landscapes, we retain the condition of variable selective pressures via a different method; namely, repeated exposures to the single, multi-modal fitness landscape (e.g., a population is repeatedly, but intermittently, exposed to a stress such as an alternate habitat, toxin, or predator population).
As per Kashtan et al.~\cite{a:kashtan:2007:01}, this creates a scenario in which evolution alternates between a ``target'' selective environment (in this case, multi-peaked) and a ``null'' selective environment where evolution is neutral.%
\footnote{%
  Kashtan et al.\ found that alternating between the target and null environments showed no speedup in most cases because there were no structural similarities between the two environments for the developmental constraints to internalize.
  In our case, however, structural similarities do exist; not between the null and target environments, but between the different local peaks of the single target environment.
  The null environment is important, however, in enabling evolution to sample multiple local peaks of the target fitness landscape.
}
Interim periods, where selection on the relevant traits is neutral, allows their values to drift and causes the next exposure to start from a different phenotype (i.e., a different location in the fitness landscape).
We term each (re-)evolution in the target landscape a microevolutionary ``episode'' and we approximate the effects of exposure to the null environment by randomizing $\symgenot$ between episodes (but retaining the slower-evolving developmental constraints, $\symbeta$).
In this manner, the slow evolution of developmental constraints responds to the phenotypic correlations selected over a distribution of different local peaks in a multi-peaked fitness landscape.
Selection occurring at multiple local peaks of multi-modal fitness landscape is conceptually similar to the notion of selection occurring in multiple, single-peaked fitness landscapes (i.e., the multiple landscapes can be viewed as different local peaks of the same underlying fitness landscape) -- but methodologically the former has the advantage that we do not need to hand-design structural similarity (e.g., modularity) into multiple phenotypic targets.

We examine two different classes of fitness landscapes each defined by the super-position of many pairwise sign-epistatic interactions.
Natural fitness landscapes contain widespread epistasis~\cite{a:weinreich:2013:01} (pairwise, or ``order two'', epistasis being the minimal case), including sign epistasis -- i.e., where a change in one trait may be either beneficial or deleterious depending on the value of another trait~\cite{a:weinreich:2005:01}.
Multiple epistatic interactions can create epistatic constraints that are difficult or impossible to resolve simultaneously (we say that an ``epistatic constraint'' between two traits is ``resolved'' if the fitness contribution conferred by their epistatic interaction is maximized; see \emph{\nameref{subsec:epistatic-fitness-landscapes}}).
Local fitness peaks occur when any change that resolves one epistatic constraint causes one or more other constraints, of greater total fitness effect, to be violated, resulting in a net decrease in fitness.
Different fitness peaks in the resultant fitness landscape tend to resolve different (but not unrelated) subsets of epistatic constraints and, in general, confer different fitness values.

A least-assumptions model of multi-modal landscapes is provided by random constraints, making no assumptions about the organization of low-order epistatic interactions.
Note that although the organization of the epistatic interactions is random, the fact that they are low-order (in this case, pairwise) means that the resultant fitness landscape, although highly multi-modal, is not arbitrary~\cite{a:watson:2011:04}.
Whether it is, in principle, possible to simultaneously resolve all epistatic constraints depends on the \emph{consistency} of the constraints in a problem~\cite{a:barahona:1982:01} -- i.e., in a consistent problem, there exists a configuration that resolves all constraints.
With random constraints, consistency is low; many epistatic constraints cannot be simultaneously resolved by \emph{any} phenotypic configuration.
Prior work with optimization using neural network learning~\cite{a:watson:2011:02} shows that the consistency of constraints will influence the success of learning (here, the ability of evolution to improve evolvability).
Accordingly, in a second landscape class, \emph{consistent constraints}, we control consistency by making all constraints consistent with a particular target phenotype.
Again this landscape exhibits many locally optimal configurations where, although there exist higher-fitness configurations, fitness improvements can require changes to many traits simultaneously.
Like the randomly organized problem, this describes a scenario where high-fitness configurations over subsets of traits are incompatible with high-fitness configurations over different but overlapping subsets of traits.
Globally optimal configurations where all constraints are resolved exist in the consistent case (but not the random case) but are nonetheless very rare and difficult to find by incremental change (in this class there are only two configurations of $2^n$ possible configurations that resolve all epistatic constraints simultaneously).

Thus our experiments simulate the evolution of a gene-regulation network on rugged fitness landscapes composed of many low-order sign-epistatic interactions.
We study the adaptation of developmental biases and constraints, accumulated over multiple evolutionary episodes that visit multiple peaks on such fitness landscapes over deep evolutionary time.
Specifically, we examine the capability of natural selection to find generalized developmental organizations that reflect the structure of these landscapes and thus modify microevolutionary trajectories in a manner that enables evolution to find different adaptive peaks of higher fitness.

\section*{Results}
In several of the experiments that follow we compare microevolutionary episodes (from random starting phenotypes): (a) with unbiased development, i.e., a one-to-one G--P map (created by using the identity matrix as the regulatory interactions matrix, $\symbeta$); (b) with an evolved developmental process, i.e., a G--P map controlled by evolved regulatory interactions (itself evolved over many past evolutionary episodes).
By comparing the end-points of these episodes we can assess whether the phenotypic distribution evolved over deep time is causing microevolutionary trajectories to find different fitness peaks in the fitness landscape and whether they are higher or lower in fitness.

\subsection*{Evolved G--P maps improve evolvability}
We begin by examining the behavior of a GRN evolving on a multi-modal fitness landscape built from either (i)~consistent epistatic constraints, or (ii)~random epistatic constraints.
Simulations for (i) and (ii) last for $1600$ and $4000$ evolutionary episodes, and each episode lasts for $5000$ and $2000$ generations respectively.

We begin by taking advantage of the fact that fit and unfit phenotypes can be easily distinguished by eye in the consistent constraints problem.
We define the problem constraints to be consistent with a recognizable image (without loss of generality, in this case, concentric squares).
Constraints are defined between neighboring pixels in the image, but finding the globally optimal phenotype requires evolution to satisfy all these local constraints simultaneously -- which is only possible by finding the target image or its complement (see \emph{\nameref{subsec:epistatic-fitness-landscapes}}).
Fitness trajectories over microevolutionary time on the landscape with consistent epistatic constraints are shown in \cref{fig:grn-vs-rhc-rc40-and-cft100-discrendfit:a}.
Here we show microevolutionary trajectories from episodes $1$--$5$, $996$--$1000$ and $1596$--$1600$.
The phenotypes found at the end of each microevolutionary episode using GRNs from different stages of deep evolutionary time are shown in the top row of \cref{fig:grn-vs-rhc-rc40-and-cft100-discrendfit:a}.
The fitnesses of phenotypes at the start of every microevolutionary episode are similar both early and late in deep evolution, but we see from the end-points of earlier and later microevolution that fitter phenotypes are found later in deep time.
Early in deep evolution, phenotypes found by microevolution have small patches that are locally consistent with the target image, but neighboring patches do not agree, leaving unresolved constraints at the boundaries between these patches.
Later in deep evolution, microevolution can find larger patches that are consistent with the target image, i.e., larger subsets of constraints are being resolved simultaneously because developmental biases have learned which combinations of traits work well together.

\begin{figure}[!tb]
  \renewcommand{\figFilename}{grn-vs-rhc-rc40-and-cft100-discrendfit}
  \centering
  \includegraphics[width=\textwidth,]{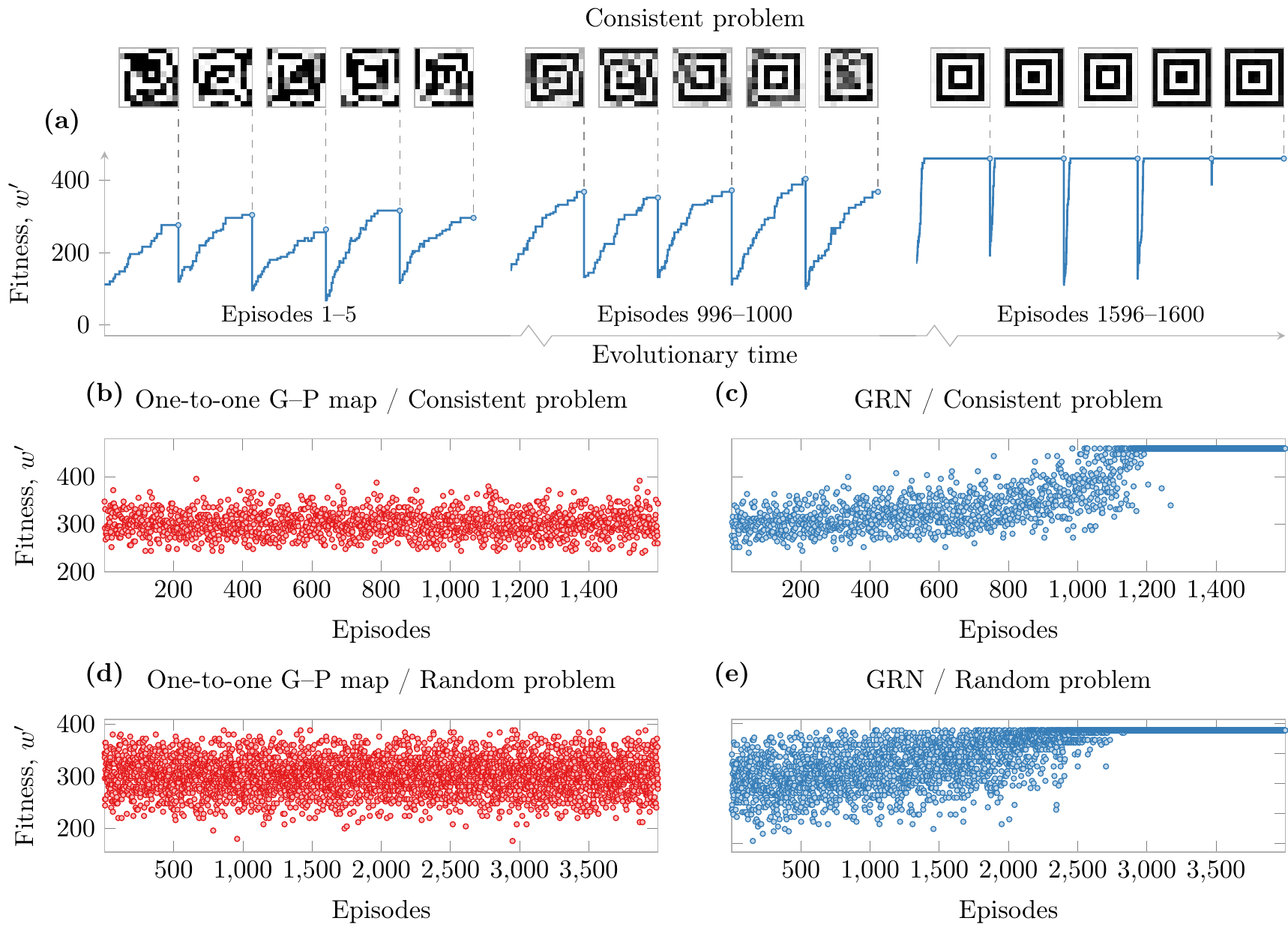}%
{%
  \ignorespaces\phantomsubcaption\label{fig:\figFilename:a}%
  \ignorespaces\phantomsubcaption\label{fig:\figFilename:b}%
  \ignorespaces\phantomsubcaption\label{fig:\figFilename:c}%
  \ignorespaces\phantomsubcaption\label{fig:\figFilename:d}%
  \ignorespaces\phantomsubcaption\label{fig:\figFilename:e}%
}
  \caption{%
    Evolution finds genotype--phenotype maps that improve evolvability.
    \subref{fig:\figFilename:a}~Fitness trajectories for the GRN over microevolutionary time on the landscape with consistent epistatic constraints.
    Top row in \subref{fig:\figFilename:a}: Evolution finds different phenotypes with every new evolutionary episode.
    With evolutionary time, evolution learns deep structural regularities common to the phenotypes it has previously encountered that allow it to increase evolvability by generalizing and successfully resolving the epistatic constraints in the fitness landscape.
    Although the fitnesses of the initial phenotypes are similar both early and late in deep evolutionary time, microevolution at later evolutionary periods finds fitter phenotypes.
    The phenotypes evolve larger patches that agree with the target image with evolutionary time.
    By the end, only the globally optimal phenotypes are found.
    Evolution over deep time on the landscape with consistent epistatic constraints is shown in \subref{fig:\figFilename:b}--\subref{fig:\figFilename:c}, and for random epistatic constraints in \subref{fig:\figFilename:d}--\subref{fig:\figFilename:e}.
    Each point in \subref{fig:\figFilename:b}--\subref{fig:\figFilename:e} shows the fitness of the phenotype at the end of a microevolutionary episode (each starting from a random $\symgenot$ phenotype).
    \subref{fig:\figFilename:b} and \subref{fig:\figFilename:d} show results for a one-to-one G--P map (red), and \subref{fig:\figFilename:c} and \subref{fig:\figFilename:e} for a GRN (blue).
    Since there is no developmental process, and thus no G--P map that can learn across deep evolutionary time, the distribution of locally optimal phenotypes found shows no trend during the whole simulation.
    In the same landscapes, with a GRN evolving slowly over deep time, evolution becomes better at discovering high-fitness phenotypes, eventually finding a very high-fitness phenotype from any starting phenotype.
    The results shown here are consistent across $30$ replicates of the experiments (\cref{fig:grn-vs-rhc-rc40-and-cft100-discrmeanendfit}).%
  }
  \label{fig:\figFilename}
\end{figure}

\begin{figure}[tb]
  \renewcommand{\figFilename}{grn-vs-rhc-rc40-and-cft100-discrmeanendfit}
  \centering
  \includegraphics[width=\textwidth,]{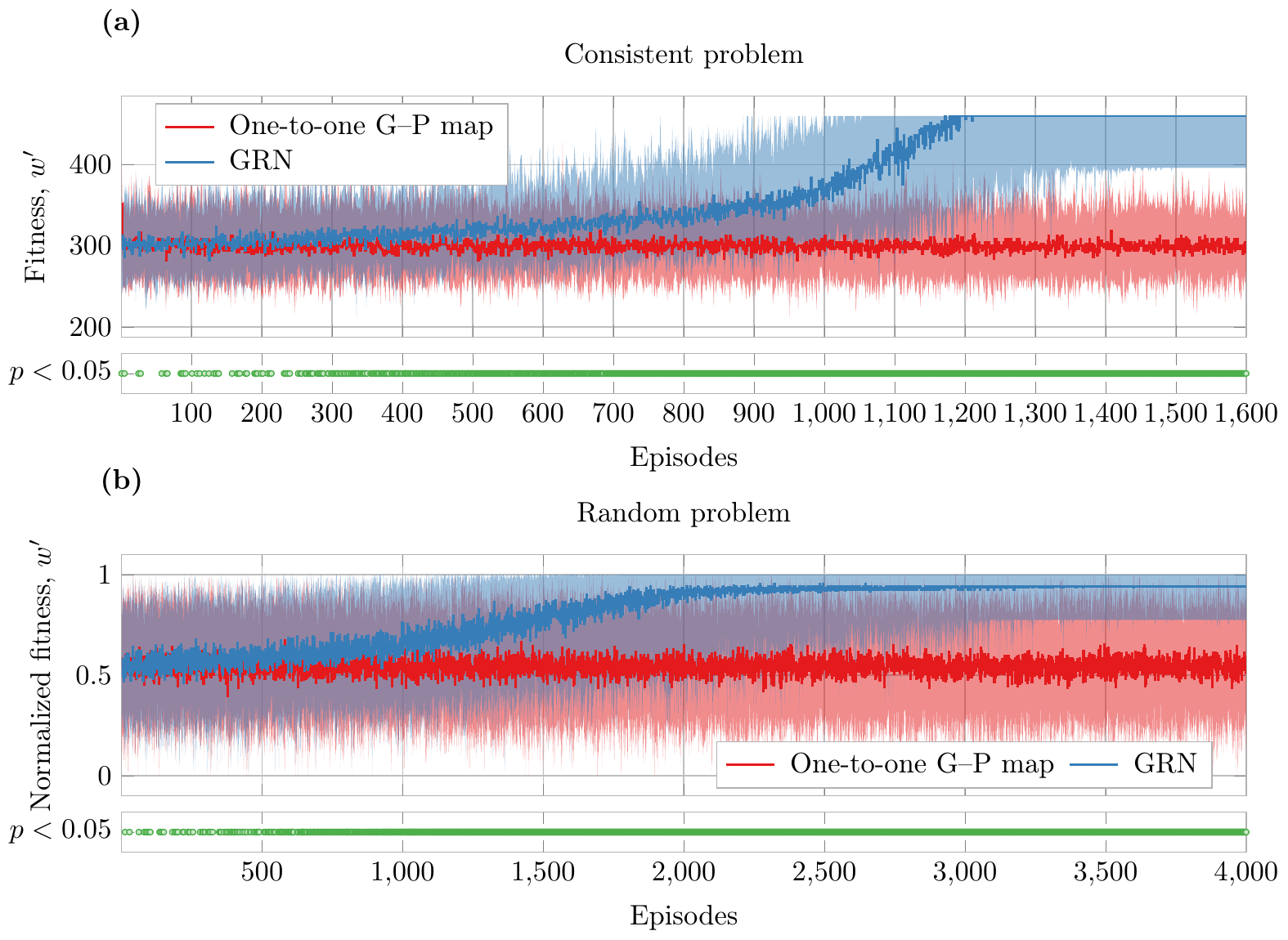}%
{%
  \ignorespaces\phantomsubcaption\label{fig:\figFilename:a}%
  \ignorespaces\phantomsubcaption\label{fig:\figFilename:b}%
}
  \caption{%
    Evolution finds G--P maps that improve evolvability.
    These plots show the median fitness of the phenotypes found at the end of every microevolutionary episode (each starting from a random phenotype) over $30$ replicates of the experiments.
    The shaded area shows the minimum and maximum fitness found by any of the $30$ replicates.
    For the random constraints experiments, we normalize fitness for each problem instance to the range $[0,1]$ with respect to the minimum and maximum fitness found by either of the two treatments.
    p-values shown are from the Mann--Whitney \emph{U} test.
    For both problem classes tested, the distribution of phenotypes found by the one-to-one G--P map shows no trend during evolutionary time.
    This is because there is no developmental process, and thus no G--P map that can learn across deep evolutionary time.
    When the GRN is evolved on these same problem classes, however, the distribution of phenotypes that evolution finds shows a trend of increasing fitness.
    With a GRN evolving slowly over deep time, evolution becomes better at discovering high-fitness phenotypes, eventually finding very high-fitness phenotypes from any starting phenotype.%
  }
  \label{fig:\figFilename}
\end{figure}

To further illustrate this progressive improvement in the ability to find high-fitness peaks, we plot the fitnesses of the phenotypes found at the end of every microevolutionary episode for one instance of the consistent and random constraints problems (\cref{fig:grn-vs-rhc-rc40-and-cft100-discrendfit:c,fig:grn-vs-rhc-rc40-and-cft100-discrendfit:e}).
We find that during the evolution of the GRN over deep time, microevolution becomes better at discovering high-fitness phenotypes.
Although the evolving GRN does not find high-fitness phenotypes reliably early in deep evolution, it slowly becomes better at evolving.
Later evolutionary episodes find better configurations by learning phenotypic correlations from locally optimal phenotypes it found in the past.
That is, evolution slowly internalizes the problem structure that can be inferred from previously-visited peaks into its G--P map, and subsequent microevolution is then biased to recreate the structures that have been learned (as shown in the top row of \cref{fig:grn-vs-rhc-rc40-and-cft100-discrendfit:a}).

To highlight the effect these evolved biases of the G--P map have on subsequent microevolution, we compare the evolved G--P map's behavior to that of a one-to-one G--P map (\cref{fig:grn-vs-rhc-rc40-and-cft100-discrendfit:b,fig:grn-vs-rhc-rc40-and-cft100-discrendfit:d} for the consistent and random constraints problems respectively).
With a one-to-one G--P map, past experience cannot alter the developmental process (there is no trend over deep time) and microevolution simply follows the fitness gradients of the fitness landscape.
Although the evolving GRN and one-to-one G--P map models find phenotypes of similar fitness early in deep evolution, the evolving GRN increases its ability to find fit phenotypes over time (\cref{fig:grn-vs-rhc-rc40-and-cft100-discrendfit:b,fig:grn-vs-rhc-rc40-and-cft100-discrendfit:c,fig:grn-vs-rhc-rc40-and-cft100-discrendfit:d,fig:grn-vs-rhc-rc40-and-cft100-discrendfit:e}).
This clearly illustrates that the evolved biases and constraints in the G--P map enable evolution to find higher-fitness phenotypes than those found by an unbiased evolutionary process (i.e., a one-to-one G--P map).
This is true for both problem classes tested, and the results are consistent over $30$ independent replicates of each experiment, with $30$ different landscapes in the random constraints case (\cref{fig:grn-vs-rhc-rc40-and-cft100-discrmeanendfit}).

\subsection*{The evolved G--P map avoids low-fitness local peaks in the fitness landscape}
We verify that the increased evolvability of the evolved GRN is due to an ability to avoid low-fitness local optima in the fitness landscape by comparing microevolutionary trajectories given (a)~a one-to-one G--P map; and (b)~G--P maps taken from different stages of the previous experiments.
We create $1000$ individuals with different random $\symgenot$ values (and hence random phenotypes) and then evolve $\symgenot$, using these different G--P maps.
To ensure that these differences are really caused by finding different local optima, not merely by finding the same phenotypes faster, we run each microevolutionary episode until phenotypes show no further improvement.

We confirm that phenotypes found with the evolved G--P maps are fitter than those found with the one-to-one G--P map, and that the longer the G--P map is evolved, the better its evolvability (\cref{fig:grn-vs-rhc-cft100-1000-genotypes-fittraj-and-endfit,fig:grn-vs-rhc-rc40-1000-genotypes-fittraj-and-endfit}).
Note that these experiments properly distinguish between individuals that are highly evolved and individuals that have high evolvability.
That is, when $\symgenot$ is randomized, phenotypes have similar fitness regardless of the G--P map.
Yet, evolution with the evolved G--P maps finds fitter phenotypes (higher end points) and it finds them faster (steeper initial slope).
Such concrete evidence for the evolution of evolvability in multi-peaked fitness landscapes has not previously been demonstrated.
We also verified that G--P maps that can achieve this level of performance are extremely rare in the space of possible G--P maps (\cref{fig:grn-cft100-fitnesses-of-phenotypes-developed-from-matrices}).
This is strong evidence that natural selection has optimized the G--P map of the evolving GRN.
Despite the fact that mutations to the G--P map are selected only on the basis of their immediate, short-term fitness consequences, the regulatory interactions that are evolved represent the problem structure, and the evolved developmental biases facilitate the production of high-fitness phenotypes.

\begin{figure}[tb]
  \renewcommand{\figFilename}{grn-vs-rhc-cft100-1000-genotypes-fittraj-and-endfit}
  \centering
  \includegraphics[width=\textwidth,]{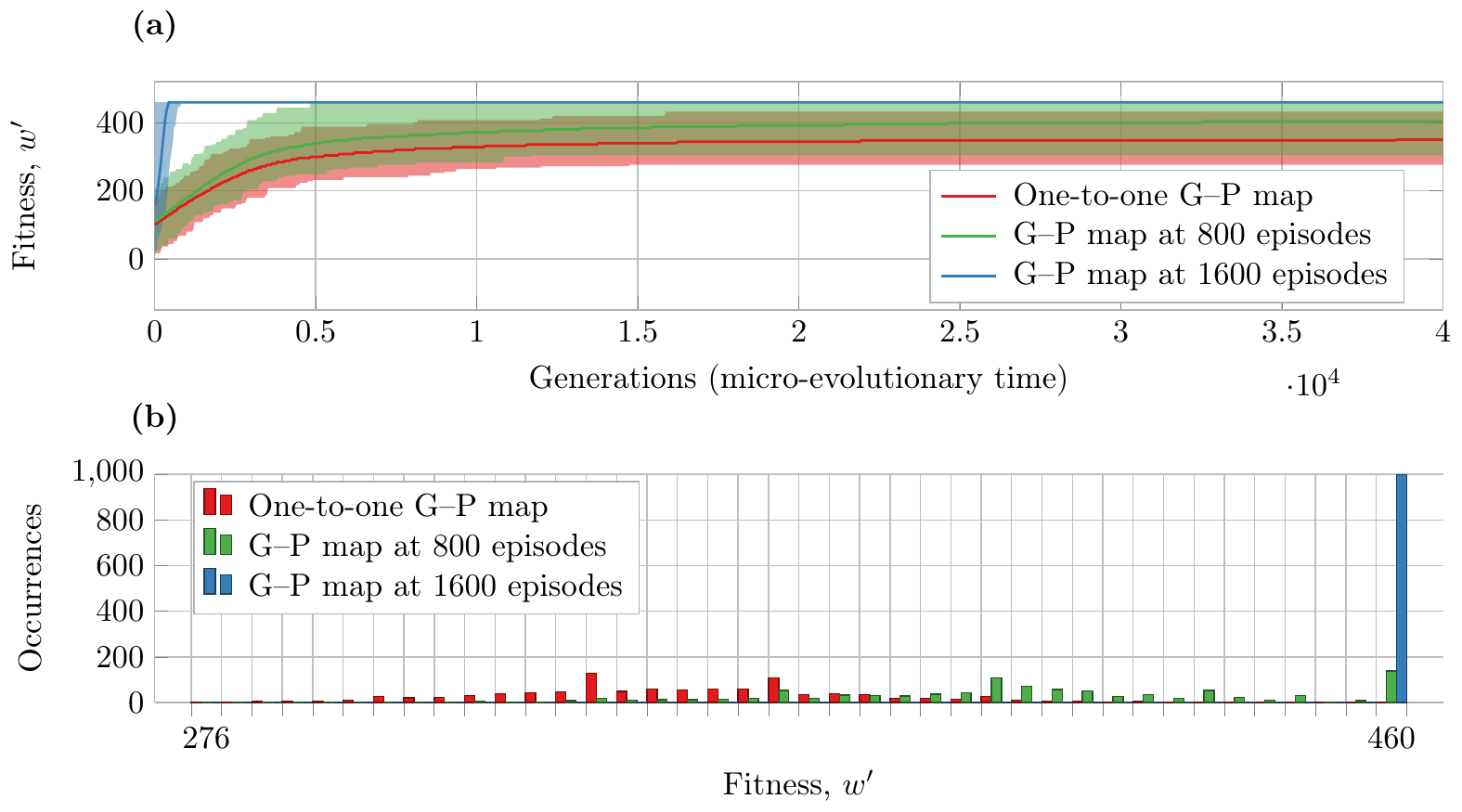}%
{%
  \ignorespaces\phantomsubcaption\label{fig:\figFilename:a}%
  \ignorespaces\phantomsubcaption\label{fig:\figFilename:b}%
}
  \caption{%
    The evolved G--P map improves the ability of microevolution to avoid low-fitness local peaks in the consistent constraints fitness landscape and find higher-fitness peaks.
    This figure shows evolution over microevolutionary time, given different G--P maps, rather than only the fitnesses found at the end of microevolutionary episodes as in \cref{fig:grn-vs-rhc-rc40-and-cft100-discrendfit}.
    We use G--P maps taken from the experiment shown in \cref{fig:grn-vs-rhc-rc40-and-cft100-discrendfit:c}, freeze their deep evolution (i.e., freeze the evolution of the G--P map $\symbeta$, but allow evolution of the genotype $\symgenot$), and probe the effect they have on subsequent microevolution.
    We randomly initialize $1000$ $\symgenot$ vectors drawn from a uniform distribution in the range $[-1,1]$ and let them microevolve for $40000$ generations (instead of $5000$ as in the previous experiment) using three G--P maps: a one-to-one mapping, and two mappings taken from episodes $800$ and $1600$ from the experiment shown in \cref{fig:grn-vs-rhc-rc40-and-cft100-discrendfit:c}.
    \subref{fig:grn-vs-rhc-cft100-1000-genotypes-fittraj-and-endfit:a}~Lines show median fitness, with the shaded areas showing the minimum and maximum fitness found by any of the $1000$ individuals for every evolutionary time-step.
    Microevolution with the less-evolved G--P maps is stuck on lower-fitness phenotypes, but microevolution with more-evolved G--P maps avoids these local peaks in the fitness landscape to find different, higher-fitness phenotypes.
    Additionally, microevolution with the more-evolved G--P maps requires less evolutionary time to find the phenotypes it settles on as can be seen from the steeper slope.
    \subref{fig:grn-vs-rhc-cft100-1000-genotypes-fittraj-and-endfit:b}~End points of \subref{fig:grn-vs-rhc-cft100-1000-genotypes-fittraj-and-endfit:a} shown in histogram format ($40$ equally sized buckets).
    Both evolved mappings find a fitter distribution of phenotypes than the one-to-one mapping.
    The evolved maps find phenotypes that are not just of higher fitness on average, but of higher fitness than \emph{all} phenotypes found with the one-to-one G--P map.%
  }
  \label{fig:\figFilename}
\end{figure}

\begin{figure}[tb]
  \renewcommand{\figFilename}{grn-vs-rhc-rc40-1000-genotypes-fittraj-and-endfit}
  \centering
  \includegraphics[width=\textwidth,]{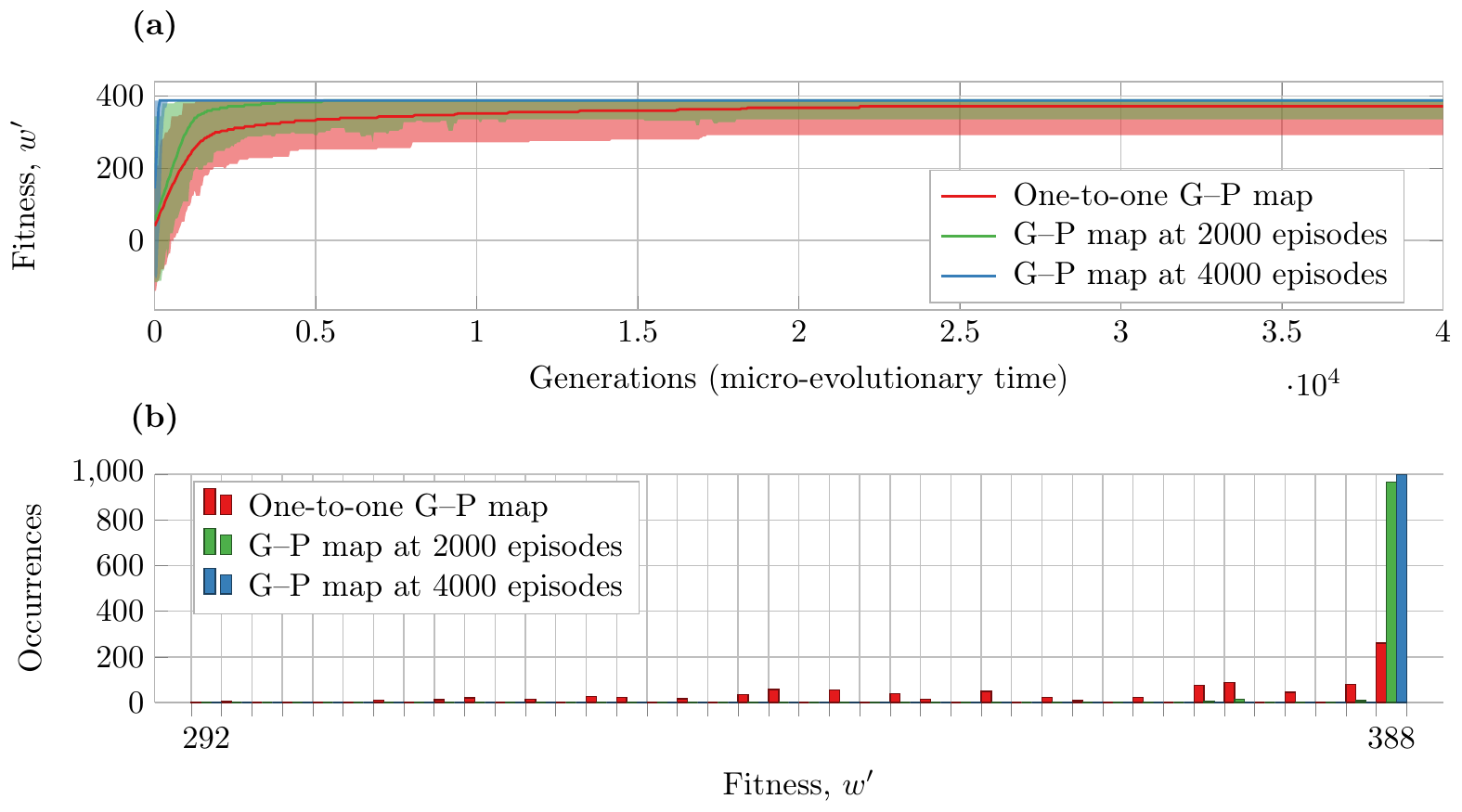}%
{%
  \ignorespaces\phantomsubcaption\label{fig:\figFilename:a}%
  \ignorespaces\phantomsubcaption\label{fig:\figFilename:b}%
}
  \caption{%
    The evolved G--P map improves the ability of microevolution to avoid low-fitness local peaks in the random constraints fitness landscape and find higher-fitness peaks.
    This figure shows evolution over microevolutionary time, given different G--P maps, rather than only the fitnesses found at the end of microevolutionary episodes as in \cref{fig:grn-vs-rhc-rc40-and-cft100-discrendfit}.
    We use G--P maps taken from the experiment shown in \cref{fig:grn-vs-rhc-rc40-and-cft100-discrendfit:e}, freeze their deep evolution (i.e., freeze the evolution of the G--P map $\symbeta$, but allow evolution of the genotype $\symgenot$), and probe the effect they have on subsequent microevolution.
    We randomly initialize $1000$ $\symgenot$ vectors drawn from a uniform distribution in the range $[-1,1]$ and let them microevolve for $40000$ generations (instead of $2500$ as in the previous experiment) using three G--P maps: a one-to-one mapping, and two mappings taken from episodes $2000$ and $4000$ from the experiment shown in \cref{fig:grn-vs-rhc-rc40-and-cft100-discrendfit:e}.
    \subref{fig:grn-vs-rhc-rc40-1000-genotypes-fittraj-and-endfit:a}~Lines show median fitness, with the shaded areas showing the minimum and maximum fitness found by any of the $1000$ individuals for every evolutionary time-step.
    Although microevolution with the less-evolved G--P maps is able to find higher-fitness phenotypes due to being allowed to microevolve for longer (compared to the phenotypes it finds in the experiment in \cref{fig:grn-vs-rhc-rc40-and-cft100-discrendfit}), microevolution with the more-evolved G--P maps avoids local peaks in the fitness landscape more reliably and finds higher-fitness phenotypes on average.
    When comparing the slope of the trajectories, we can see that microevolution with the more-evolved G--P maps settles on high-fitness phenotypes faster than microevolution with the less-evolved G--P maps.
    \subref{fig:grn-vs-rhc-rc40-1000-genotypes-fittraj-and-endfit:b}~End points of \subref{fig:grn-vs-rhc-rc40-1000-genotypes-fittraj-and-endfit:a} shown in histogram format ($40$ equally-sized buckets).
    Both evolved mappings find a fitter distribution of phenotypes than the one-to-one mapping.
    The evolved maps find the best phenotypes found by the one-to-one G--P map more reliably.%
  }
  \label{fig:\figFilename}
\end{figure}

\begin{figure}[tb]
  \renewcommand{\figFilename}{grn-cft100-fitnesses-of-phenotypes-developed-from-matrices}
  \centering
  \includegraphics[width=\textwidth,]{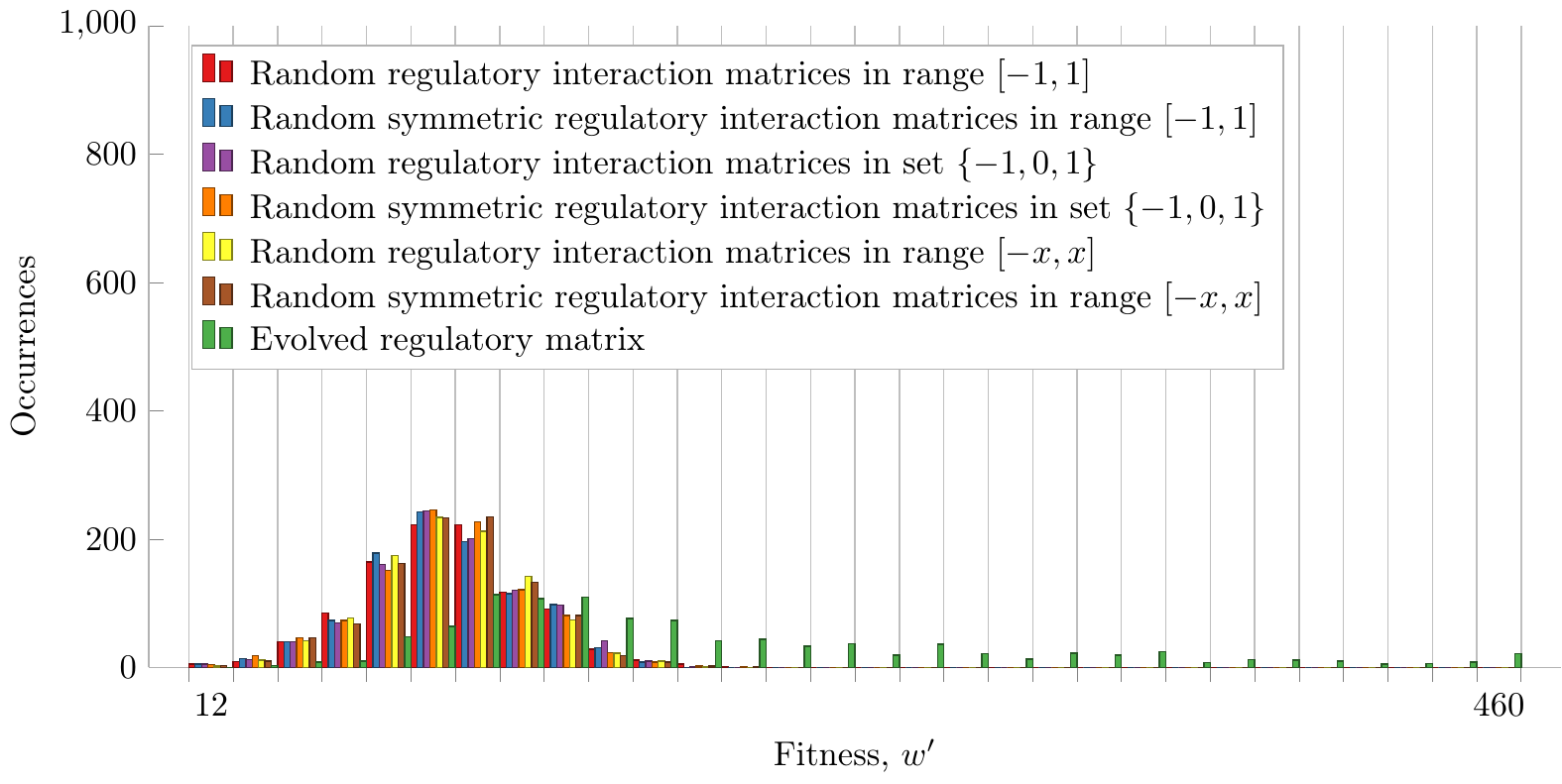}%
  \caption{%
    Natural selection has optimized the G--P map for long-term evolvability.
    We randomly initialize $1000$ $\symgenot$ vectors drawn from a uniform distribution in the range $[-1,1]$ and evaluate the adult phenotypes they develop into using different regulatory interaction matrices.
    For each $\symgenot$ vector, we randomly initialize a regulatory interaction matrix $\symbeta$ in different ranges and sets (see legend; $x$ is the mean value of all magnitudes in the final, evolved matrix coming from \cref{fig:grn-vs-rhc-rc40-and-cft100-discrendfit:c}), and evaluate the phenotype $\symgenot$ is developed into given the interaction matrix $\symbeta$.
    Since selective pressures in the fitness landscape are symmetric and our evolved matrices are also approximately symmetric, we force some of the regulatory interaction matrices we generate to be symmetric.
    For comparison, we evaluate the adult phenotypes that are produced from the same $1000$ $\symgenot$ vectors, but using the regulatory interaction matrix taken from the end of the experiment shown in \cref{fig:grn-vs-rhc-rc40-and-cft100-discrendfit:c}.
    We find no significant difference in performance between symmetric and non-symmetric randomly generated matrices.
    On the other hand, the evolved G--P map develops phenotypes of higher fitness on average.
    This is because natural selection optimized the G--P map of the evolving GRN.
    Even though it was selected for immediate, short-term fitness advantages, the regulatory interactions that are evolved represent the problem structure, and the evolved developmental biases facilitate the production of high-fitness phenotypes.
  }
  \label{fig:\figFilename}
\end{figure}

\subsection*{The evolved G--P maps internalize structural information about the selective environment}
To confirm that the evolved G--P map is internalizing structural information about the selective environment, as Riedl suggested~\cite{a:riedl:1977:01,b:riedl:1978:01}, we analyze the structure of the evolved regulatory interactions after evolving the GRN on the consistent problem.
The structural information within the selective environment can be seen in terms of the epistatic interactions defined in the problem matrix (\cref{fig:grn-cft100-imitatory-epigenotype}, first frame; white and black for $+$ and $-$ epistatic interactions between gene $i$ and gene $j$ respectively, grey for no interactions).
In episode $1$, the regulatory interactions evolved reflect the traits in the one local optimum discovered at the end of that episode.
In episode $2$, a different local optimum was visited, so conflicting regulatory interactions were evolved.
Thus, interactions which existed in both phenotypes were reinforced (black and white), whereas interactions which existed only in one phenotype were weakened (grey).
With evolutionary time, after a number of different local optima have been visited, regulatory interactions closely match the structural information which exists in the selective environment (episodes $100$ and $800$).
Both positive and negative regulatory interactions have evolved along the diagonals, but the rest of the regulatory interactions remain close to zero, as is the case in the problem matrix.
At this stage, the developmental architecture has successfully internalized structural information about the selective environment.

\begin{figure}[tb]
  \renewcommand{\figFilename}{grn-cft100-imitatory-epigenotype}
  \centering
  \includegraphics[width=\textwidth,]{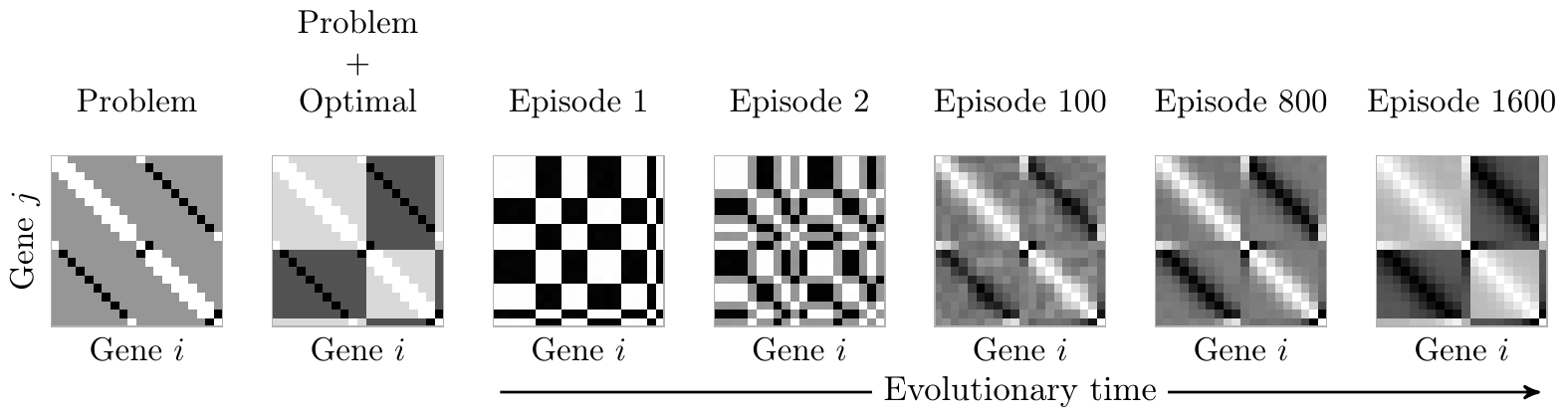}%
  \caption{%
    The evolving G--P map goes beyond internalizing structural information about the selective environment, finding a system of constraints that enables evolution to find the global optima reliably.
    The first frame shows a subset ($20\times20$ out of $100\times100$) of the consistent problem matrix ($\symvect{C}$; see \emph{\nameref{subsec:epistatic-fitness-landscapes}}), and frames $3$--$7$ show visual representations of a subset of the regulatory interaction matrices at different stages of deep evolution (also $20\times20$ subsets; white for maximally positive numbers at that episode, black for maximally negative numbers and greyscale for intermediate values).
    The second frame shows a $20\times20$ subset of the regulatory interaction matrix we approximate evolution will evolve due to visiting a large number of local optima and not only the global optima (see text).
    This set of regulatory interaction enables evolution to find the global optima reliably.
    The evolved G--P maps at episodes $1$ and $2$ are consistent with the local optima discovered at the end of each respective episode, but differences between the two phenotypes cause evolution to reinforce interactions which exist in both phenotypes (black and white) while weakening interactions which exist in only one (grey).
    After a number of local optima have been visited, the evolved regulatory interactions (episodes $100$ and $800$) closely resemble the problem matrix (first frame).
    Both positive and negative regulatory interactions have evolved along the diagonals (darker and lighter colors), but the rest of the regulatory interactions remain close to $0$, as is the case in the problem matrix.
    At this point, the evolved G--P maps have internalized structural information about the selective environment -- i.e., what values patches of neighboring pixels should take to confer fitness benefits, but not how these patches should be combined at the boundaries to produce the globally optimal phenotypes.
    This developmental architecture serves as an intermediate phase between learning the problem structure and finding a system of constraints that enables evolution to find the global optima reliably.
    At the end of the simulation (episode $1600$), the evolved regulatory interaction matrix closely resembles this approximated optimal matrix (second frame).
    As is the case earlier in evolution, both positive and negative regulatory interactions have evolved along the diagonals (darker and lighter colors).
    However, the rest of the regulatory interactions don't remain close to zero, but instead evolve to up-regulate or down-regulate in a way which is consistent with the globally optimal phenotypes.%
  }
  \label{fig:\figFilename}
\end{figure}

We find that this developmental architecture serves as an intermediate phase between learning the problem structure and finding a system of constraints that enables evolution to find the global optimum reliably.
Since we know what the globally optimal phenotype is in the consistent problem, we also know what the optimal regulatory interaction matrix is: the matrix whose regulatory interactions match the pixel correlations in the concentric squares pattern.
In a multi-modal fitness landscape, such as the one we use here, the phenotype found by evolution at the end of every microevolutionary episode is, in practically every case, not the global optimum.
However, the different local optima found at the end of every microevolutionary episode provide evolution with different, partially contradictory and partially consistent, information about combinations of phenotypic traits that are fit.
Given sufficiently many examples, the evolving GRN begins to average-out the contradictory information and amplify the consistent information.
Therefore, we hypothesize that the evolved regulatory interaction matrix will, in the medium term, become similar to the problem matrix -- because the constraints in the problem matrix reveal themselves in the fitness landscape as the most consistent ``signal'' that evolution can identify (this is the imitatory organization that Riedl described).
But later, we would expect the evolved GRN to ``fill-in'' entries in this matrix that were unspecified in the problem matrix -- and eventually, to identify values that correspond to the optimal solution matrix (\cref{fig:grn-cft100-imitatory-epigenotype}, last frame).
Although the global optimum might not yet have been visited, this filling-in of values that agree with the global optimum is nonetheless possible because these interactions are the ones that are most reliably consistent with the correlations observed over a distribution of local optima~\cite{a:watson:2014:01}.

Our data confirms that this behavior, previously studied in a learning neural network~\cite{a:watson:2011:02}, also occurs in the evolving GRN.
At the end of the simulation, the evolved regulatory interaction matrix closely resembles the addition of the problem and optimal matrices as described above (\cref{fig:grn-cft100-imitatory-epigenotype}; second and last frames).
A clear separation between positive and negative regulatory interactions persists along the diagonals, and the rest of the regulatory interactions also evolve to up-regulate or down-regulate in a way which is consistent with the globally optimal phenotype.
The evolving G--P map thus goes beyond internalizing structural information about the selective environment, finding a system of constraints that enables evolution to find the global optima reliably.

\subsection*{Evolvability evolves as a systematic by-product of short-term selection for fitness}
Here we begin to unpack how the evolution of evolvability occurs in these experiments.
Since the long-term fitness consequences cannot be the reason that natural selection evolves these highly evolvable G--P maps, we need to identify the immediate selective pressures in operation and how they have this effect on long-term evolvability.

To understand how short-term selective pressures affect long-term evolvability, we set $\symgenot$ to match the concentric squares image and evolve only the regulatory interactions $\symbeta$.
We examine how the fitness of the adult phenotype produced changes as the regulatory interactions evolve.
Note that $\symgenot$ has to develop according to the regulatory interactions, $\symbeta$, and these interactions are initially absent and slowly evolve.
Because of that, the evolution of $\symbeta$ has a fitness effect even though $\symgenot$ already matches the concentric squares image.
We also examine how the developmental basin of attraction of the concentric squares phenotype changes -- i.e., the number of $\symgenot$ vectors that map into the concentric squares phenotype.

When GRN interactions evolve they increase the fitness of the phenotype; otherwise they would not be selected by short-term selection.
But they also have an effect on the distribution of phenotypes produced by that GRN under subsequent mutations to $\symgenot$.
Specifically, both the fitness of the adult phenotype and the developmental basin of attraction of the concentric squares phenotype increase with evolutionary time (\cref{fig:grn-cft100-s10000-fitness-and-dboa}).
That is, regulatory interactions that increase immediate fitness necessarily have the systematic side-effect of increasing the number of genotypes that map into that same phenotype.
This is because the evolved regulatory interactions reinforce correlations which exist in the current phenotype and these correlations bias the distribution of phenotypes that can be produced by the developmental process towards reproducing these same evolved correlations~\cite{a:watson:2014:01}.
In this sense, we say that the genetic robustness of the selected phenotype is increased -- the number of $\symgenot$ vectors that map into a specific phenotype is increased.
This increase in the developmental basin of attraction of the phenotype is functionally equivalent to the development of an associative memory for the current phenotype~\cite{a:watson:2014:01}.

\begin{figure}[tb]
  \renewcommand{\figFilename}{grn-cft100-s10000-fitness-and-dboa}
  \centering
  \includegraphics[width=\textwidth,]{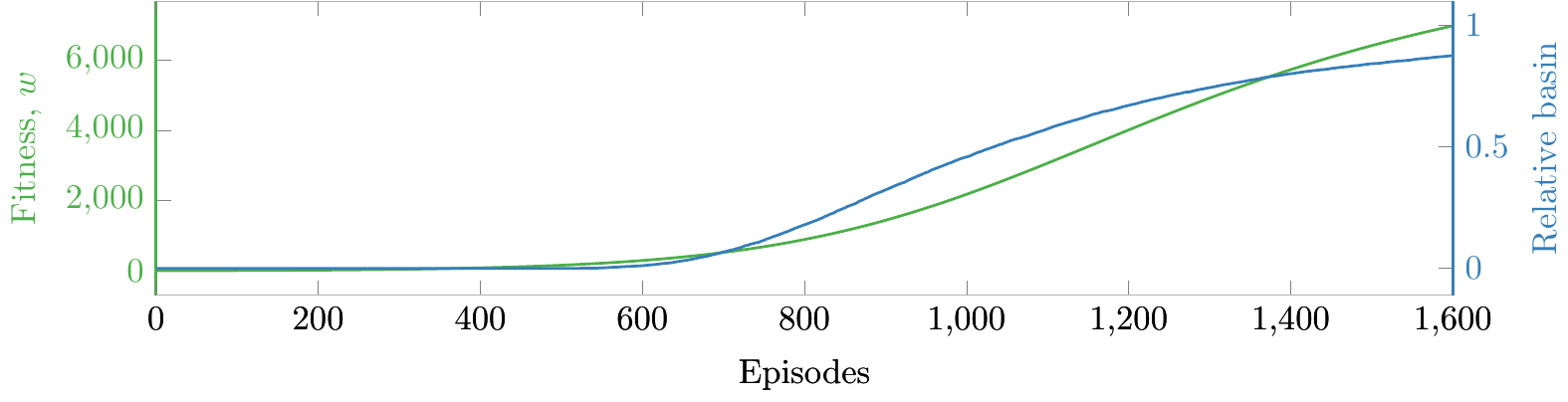}%
  \caption{%
    Selection for fitness increases the developmental basin of attraction for the target phenotype.
    Without loss of generality, we make $\symgenot$ equal to the target phenotype of the consistent constraints problem, and evolve \emph{only} the regulatory interactions under natural selection for $1600$ evolutionary episodes.
    We measure how the fitness of phenotypes developed from $\symgenot$ changes with evolutionary time and also how the developmental basin of attraction of the target phenotype changes with time.
    The fitness of the adult phenotype produced increases with evolutionary time (green; left y-axis) due to the developmental interactions evolving in the direction of the target phenotype, thus amplifying the adult phenotype's gene expression levels (this figure does not use fitness of discretized phenotypes; see \emph{\nameref{subsec:epistatic-fitness-landscapes}}).
    The proportion of the developmental basin of attraction of the target phenotype also changes with evolutionary time.
    This is measured by randomly sampling the $\symgenot$ space (here, sample size is set to $10000$ in a uniform distribution in the range $[-1,1]$), developing all $\symgenot$ vectors using G--P maps taken from the end of every evolutionary episode, and reporting the proportion of $\symgenot$ vectors that develop into the target phenotype.
    The basin of attraction of the target phenotype increases with evolutionary time (blue; right y-axis).
    The target phenotype's basin of attraction does not see an immediate increase because earlier in evolution the regulatory interactions are too ``weak'' to solely determine the adult phenotype produced.
    As the regulatory interactions become stronger with evolutionary time, the basin of attraction increases.
    By the end of the simulation, the basin of attraction of the target phenotype enlarges to encompass almost the entire space, such that $\approx90\%$ of the sampled $\symgenot$ vectors develop into the target phenotype.
  }
  \label{fig:\figFilename}
\end{figure}

This robustness does not, in itself, explain the evolution of evolvability -- indeed, an inability to produce different phenotypes opposes evolvability.
But what happens to developmental basin sizes when regulatory interactions evolve slowly over multiple selected phenotypes as in our main experiments?

\begin{figure}[tb]
  \renewcommand{\figFilename}{grn-cft100-cross-section-with-evolutionary-time}
  \centering
  \includegraphics[]{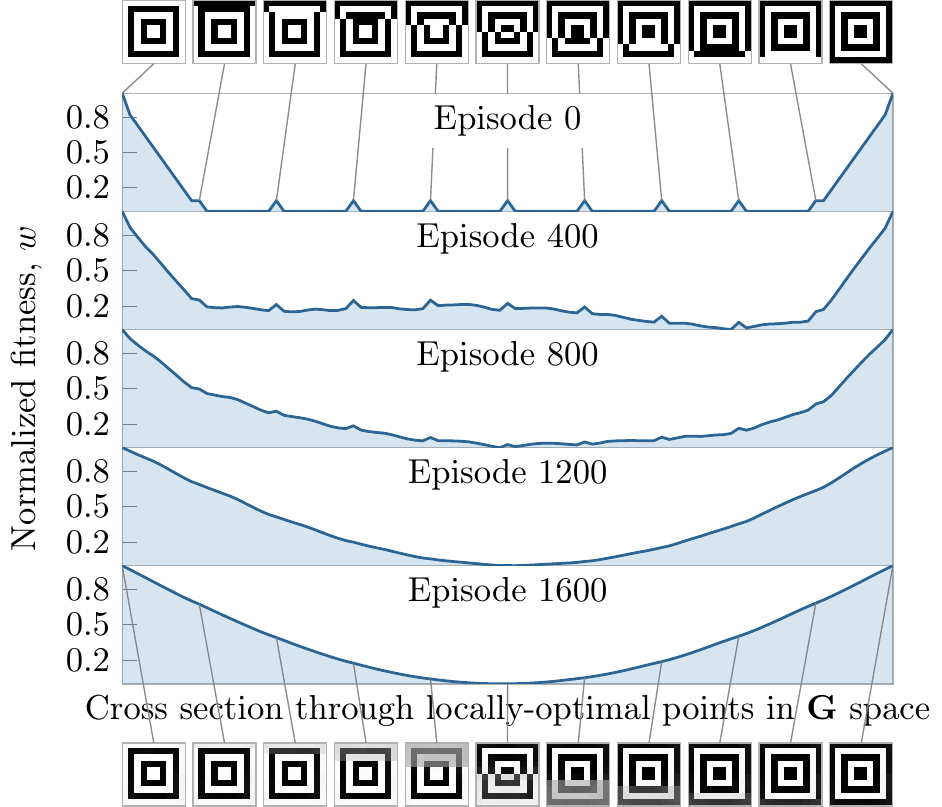}%
  \caption{%
    The evolving G--P map enhances evolvability by enlarging the basins of the global optima and removing basins for local optima.
    We evolve a GRN on the consistent constraints problem for $1600$ episodes, and store all G--P maps from the end of every episode.
    We then use a cross section through the space of $\symgenot$ values that starts from the global optimum (concentric squares) and passes through a number of local optima in the fitness landscape.
    The genotypes in this cross-section are developed into adult phenotypes using all $1600$ G--P maps from the previous simulation, and the adult phenotypes are evaluated.
    The fitnesses of phenotypes developed with each G--P map, in each cross-section, are normalized for comparison.
    In the case of a one-to-one map (first frame; episode $0$), the adult phenotypes produced are equal to the $\symgenot$ vectors, and the fitness landscape is multi-modal, meaning that evolution can get trapped on local optima.
    The top row shows the locally optimal adult phenotypes produced.
    Over many evolutionary episodes the GRN changes.
    This does not change the mapping from phenotypes to fitness, but it does change the mapping from genotypes ($\symgenot$) to phenotypes, as evident from the differences in the phenotypes depicted at the top and bottom rows.
    Accordingly, the fitness landscape for this cross-section through $\symgenot$ changes over evolutionary time.
    Specifically, the fitness landscape produced by the phenotypes developed from this cross-section of $\symgenot$ becomes progressively smoother, eliminating local optima and enlarging the basin of attraction of the global optimum.
    Thus the fitness landscape available to selection slowly becomes effectively uni-modal when augmented by the evolving G--P map.%
  }
  \label{fig:\figFilename}
\end{figure}

Given that the sum of all basins of attraction has a fixed size (the total configuration space of $\symgenot$), an increase in one basin implies a decrease in the combined size of the others.
Thus, when a GRN is evolved over a distribution of phenotypes, not all of their basins can increase indefinitely and, to first approximation, the phenotypes that are selected for most often will eventually out-compete all other phenotypes.
However, evolving interactions store information about phenotypic correlations rather than the specific values of individual phenotypic traits~\cite{a:watson:2014:01}.
Accordingly, when one basin increases others will decrease, but the basins of phenotypes that are structurally similar to the currently selected phenotype will reduce only a little compared to the basins of phenotypes that are structurally unrelated.
Over a distribution of selected phenotypes, the GRN learns a generalized model of those phenotypes -- i.e., it learns their correlation structure -- and the phenotypes whose basins are enlarged the most are those that agree with the most phenotypic correlations observed over that sample of selected phenotypes (see ref.~\cite{a:watson:2011:02} for an illustration of this in neural network learning).

We use the consistent problem to confirm that as the GRN evolves, the size of the basin of local optima decreases and the size of the basin of the global optimum increases even though the global optimum has not yet been visited (\cref{fig:grn-cft100-cross-section-with-evolutionary-time}).
This, combined with the fact that evolved regulatory interactions are consistent with the globally optimal phenotype (\cref{fig:grn-cft100-imitatory-epigenotype}), explains how microevolutionary trajectories are able to follow fitness gradients and find the global optimum even though phenotypes of similar fitness (i.e., starting in the same place) with a one-to-one or less-evolved developmental G--P may have instead followed fitness gradients to a local optimum (as depicted in \cref{fig:grn-cft100-cross-section-with-evolutionary-time-with-trajectories,fig:grn-vs-rhc-rc40-and-cft100-discrendfit:a}).

\section*{Discussion}
Our results demonstrate conditions where short-term selection adapts developmental constraints such that they facilitate long-term adaptation on rugged fitness landscapes.
This is possible when the landscape is composed of low-order epistatic interactions because this presents structural regularities (i.e., phenotypic correlations) that a GRN can represent and can acquire through selection.
We show that evolvability is an indirect but necessary consequence of selection for immediate, short-term fitness increases.
Robustness of the currently-selected phenotype is also a side effect of such changes.
But counter to naive intuitions, robustness does not oppose evolvability -- in fact, canalization (of the right things) is necessary for evolvability~\cite{b:riedl:1978:01,a:kirschner:1998:01,a:gerhart:2007:01,a:kouvaris:2015:01}.
In this work we show that a GRN can evolve to favor fit phenotypic correlations, rather than merely favoring individual phenotypic traits, if it is exposed to selection that is representative of fit correlations over evolutionary time.
And we have shown that this is the case when a GRN is repeatedly exposed to a multi-modal fitness landscape causing it to visit a distribution of locally optimal phenotypes.
We speculate that negative frequency dependence, causing a population to continually explore phenotype space~\cite{ic:watson:2014:01}, may produce similar results to the model of repeated exposures to an otherwise static stress environment used here.

As evolved developmental biases and constraints internalize the structure of the fitness landscape, the distribution of phenotypic variation produced by development changes to favor these correlations (enriches for the production of phenotypes that contain these correlations).
This has the consequence that high-fitness phenotypes are found faster and with greater probability.
Although natural selection is myopic, selective pressures under these conditions result in a G--P map that acts like a longer-term memory and allows subsequent evolution to effectively ``ignore'' locally optimal configurations that could potentially trap evolution with an unbiased G--P map (Fig.~\ref{fig:grn-cft100-cross-section-with-evolutionary-time-with-trajectories}).
Thus, our results show the evolution of long-term evolvability as a consequence of selection for immediate fitness differences.

These behaviors can be explained by simple learning concepts.
If the past is a good indicator of the future, learning from the past can improve a model's ability to perform well in environments not previously experienced.
This requires, however, that natural selection has an appropriately powerful ``model space'', capable of representing the way in which phenotypes that were fit in the past are ``similar'' to novel high-fitness phenotypes~\cite{a:watson:2016:01}.
A one-to-one G--P cannot do this unless the similarity is simply based on Euclidean distance in phenotype space.
But an evolving GRN is capable of internalizing information about phenotypic correlations.
When this is a good model of the structure present in the selective environment, as it is when the environment contains low-order epistasis, this results in improved evolvability.
More generally, the relationship between the evolution of evolvability and generalization in learning systems helps us understand the conditions for, and limitations of, the evolution of evolvability~\cite{a:kouvaris:2015:01}.
This has enabled us to demonstrate the first formal model for the evolution of evolvability capable of showing and explaining the evolution of evolvability that facilitates future innovation from short term selection pressures.
Contrary to popular assertions, this is not impossible -- just as predicting the future by generalizing from past experience is not impossible in learning systems.
This is an example of how the transfer of machine learning theory to evolutionary theory~\cite{a:watson:2014:01,a:power:2015:01,a:kouvaris:2015:01,a:watson:2016:01,a:watson:2016:02} has potential to demystify the evolution of evolvability, and help us understand how it works and the conditions that enable it.

\section*{Materials and Methods} \label{sec:materials-and-methods}
\subsection*{Gene-regulation network} \label{subsec:gene-regulation-network}
We simulate the evolution of a gene-regulation network using a non-linear, recurrent developmental process, as described in our earlier work~\cite{a:watson:2014:01}.

The phenotype of an individual is a gene expression profile, $\symphenot$, described by a set of $n$ gene expression potentials which are assumed to control phenotypic traits, naturally represented by a vector: $\symphenot = \langle P_1, P_2, \dotsc, P_n \rangle, P \in \mathbb{R}^n$.
The genotype of an individual consists of two parts: a vector, $\symgenot = \langle G_1, G_2, \dotsc, G_n \rangle$, defining the gene expression potentials of the embryonic phenotype, and the elements, $b_\mathit{ij}$, of an interaction matrix, $\symbeta$ (a similar setup was used in refs.~\cite{a:lande:1983:01,a:lipson:2002:01,a:jones:2007:01,a:kashtan:2009:01,a:watson:2014:01}).
Every element of the interaction matrix, $\symbeta$, represents the magnitude and sign (excitatory or inhibitory) of the interaction between two traits -- that is, whether a gene up-regulates or down-regulates another gene.

We use a non-linear, recurrent developmental process that maps a genotype to a phenotype, modeled as follows.
At developmental time-step $t = 0$, the embryonic phenotype is set to $\symphenot(0) = \symgenot$.
For every subsequent developmental time-step, the phenotype vector is updated by a non-linear transformation determined by the weighted effect of each trait (or expression potential) on each other trait, (the interaction matrix) $\symbeta$, and a degradation rate:
\begin{equation}
  \symphenot(t+1) = \symphenot(t) + \sigma\left(\symbeta \times \symphenot(t)\right) - \tau\symphenot(t),
\end{equation}
where $\tau = 0.2$ is the rate of degradation, and $\sigma$ is a sigmoidal function (applied to all elements of the phenotype vector) that non-linearly limits the effect of interactions.
The non-linearity is modeled with the hyperbolic tangent: $\sigma(x) = \tanh(x)$.

The gene expression profile of the adult phenotype determines the fitness of an organism.
We define ``adult phenotype'', $\symadultphenot$, to be the phenotype vector after a fixed number of developmental time-steps, $\symdevotime = 10$, has passed: $\symadultphenot = \symphenot_{\symdevotime}$.

\subsection*{Evolutionary model} \label{subsec:evolutionary-model}
All elements of both parts of the genotype, $\symgenot$ and $\symbeta$, are initialized to $0$ (i.e, no regulatory interactions).
Every generation, mutations are applied to $\symgenot$.
Specifically, a single randomly chosen trait of $\symgenot$ is mutated every evolutionary time-step by adding to it $\mu_1$, drawn from a uniform distribution in the range $\pm0.1$.
All elements of $\symgenot$ are hard-bounded in the range $[-1,1]$.
Mutation and selection on $\symbeta$ is approximated with a selection-limited model of evolution applied at the end of every evolutionary episode (see \emph{\nameref{subsec:selection-limited-evolution}}).

Our model operates under ``strong selection, weak mutation'' assumptions~\cite{a:gillespie:1984:01}, which means that a mutation is fixed or lost in the population before another mutation occurs.
To accommodate these assumptions, the population mean genotype is represented by a single genotype ($\bar{\symgenot}$ and $\bar{\symbeta}$), which is mapped into the population mean adult phenotype, $\sympmadultphenot$ via the developmental process.
The evolution of the population is modeled by introducing small mutations to the genotype ($\bar{\symgenot}$ and $\bar{\symbeta}$), which yields a mutant genotype ($\bar{\symgenot}'$ and $\bar{\symbeta}'$).
The mutant genotype is developed into a mutant adult phenotype $\sympmadultphenot'$.
The probability of the mutant fixing in the population is proportional to the selective advantage it confers, but here we are interested in the direction of selection rather than the magnitude.
As such, we assume a ``hill-climbing'' model of selection (e.g., refs.~\cite{a:kashtan:2009:01,a:watson:2014:01}), meaning that the selection coefficient is assumed to be large enough to cause beneficial and deleterious mutations to fix or not fix respectively.
Under these assumptions, the mutant genotype becomes the mean population genotype if the fitness of its corresponding phenotype is greater than the fitness of the existing mean population phenotype ($\bar{\symgenot}(t+1) = \bar{\symgenot}'$, $\bar{\symbeta}(t+1) = \bar{\symbeta}'$), otherwise the genotype remains the same for the next generation.

\subsection*{Selection-limited evolution} \label{subsec:selection-limited-evolution}
Under \emph{directional} selection, the cumulative effect of a large number of small mutations is equivalent to the effect of a small number of large mutations when controlling for variance.
Assuming that the number of generations required for $\symgenot$ to align itself with a local optimum and stabilize is small compared the total number of generations within an evolutionary episode, the effect natural selection has in any evolutionary episode can be modeled as follows.
Hill-climbing selection is applied to all elements in $\symbeta$ (random order; no replacement) by testing two mutations: one for positive directional selection drawn from a distribution with mean $q$ and one for negative directional selection drawn from a distribution of $-q$, each with a standard deviation of $\sigma$.
Only one mutation can be selected for under directional selection.
Experiments use $q=3\times10^{-6}$ and $q=2\times10^{-6}$ for the consistent and random constraints problem classes respectively.
A standard deviation of $\sigma=0.01q$ is used for both problem classes.

With a standard deviation of $1\%$, we are simulating the cumulative effect of $10^4$ discrete mutations (each with equal probability of being accepted given directional selection) all occurring before each new evolutionary episode (coming from $\sigma = \sqrt{np(1 - p)}, (p = 0.5)$).
Note that there is a trade-off between (a) the number of mutations that occur before each new evolutionary episode; and (b) the number of evolutionary episodes that are simulated in total.
If we increase the number of evolutionary episodes, the number of mutations can be decreased, but this means that more mutation--selection cycles must be simulated.
For our purposes, we find that $1600$ evolutionary episodes are sufficient for the consistent constraints problem, and $4000$ evolutionary episodes for the random constraints problem.
This combination of control parameters enables the discovery of high-fitness phenotypes without ``over-fitting'' to a locally optimal, low-fitness peak~\cite{a:kouvaris:2015:01}.

\subsection*{Epistatic fitness landscapes} \label{subsec:epistatic-fitness-landscapes}
We utilize multi-modal selective environments defined by the sum of a large number of pairwise sign-epistatic interactions.
The fitness, $w$, of a phenotype, $\symadultphenot$, is defined by a matrix of epistatic interactions, $\symvect{C}$, as follows:
\begin{equation} \label{eqn:fitness-function}
  w(\symadultphenot) = \symadultphenot\symvect{C}\symvect{P}_a^{\mathsf{T}}.
\end{equation}

$\symvect{C}$ is an $n \times n$ symmetric matrix whose elements, $c_\mathit{ij}$, determine the nature of the epistatic interaction between traits $i$ and $j$.
All features of the fitness landscape are thus defined by the epistatic constraints defined by $\symvect{C}$.
We say that the ``epistatic constraint'' between two traits is ``resolved'' if and only if the fitness contribution from their interaction is maximized.
This is easily achieved for a single epistatic interaction (either by ``$++$'' or ``$--$'' for positive epistasis, i.e., $c_\mathit{ij} > 0$, or by ``$+-$'' or ``$-+$'' for negative epistasis, i.e., $c_\mathit{ij} < 0$, where $+$ and $-$ represent the maximal positive and negative traits values respectively), but multiple epistatic interactions between many traits can create epistatic constraints that are difficult or impossible to resolve simultaneously.

We utilize two different classes of such fitness landscapes.
Random constraints uses a problem matrix whose diagonal elements take the value $1$, $c_\mathit{ii} = 1 \forall i$, and the rest are uniformly initialized to a fixed-magnitude, variable-sign value: $c_\mathit{ij} \in \{-d, d\}$.
Here, we use $d = 1$ and set the problem size to $n = 40$.
This results in a problem with low consistency, meaning that, for example, there exists no phenotype that resolves the constraints between $A$ and $B$, and $B$ and $C$, and also resolves the constraint between $C$ and $A$.
In this fitness landscape, the globally optimal phenotype tends not to be able to resolve all epistatic constraints.

In the consistent constraints problem class, the global optimum resolves all constraints.
Here, $\symvect{C}$ is defined from a given target phenotype in a way that makes the target phenotype and its complement the globally optimal phenotypes.
Without loss of generality, we choose the target phenotype to be a $10 \times 10$, black-and-white image of concentric squares.
In doing so, we are able to provide an easily-interpretable visual representation of phenotypes (see Fig.~\ref{fig:grn-vs-rhc-rc40-and-cft100-discrendfit}\emph{a}).
The problem matrix is generated by calculating the outer product of the flattened image vector, $\symvect{x}, x_i \in \{-1, 1\}$, and setting all elements of the resulting matrix that are non-neighboring pixels with respect to the target image to zero.
Because the problem matrix contains information about neighboring pixels only (since all non-neighboring pixels are set to $0$), local optima are caused by conflicting constraints arising when local patches of constraints are resolved in mutually incompatible ways, leaving unresolved conflicts at the boundaries between these patches.

Whenever the sign of a trait resolves more constraints than it violates, \cref{eqn:fitness-function} defines directional selection that rewards increases in the trait magnitude.
This creates a sustained selective pressure for genetic mutations that alter the G--P map~\cite{a:pavlicev:2011:01,a:watson:2011:02}.
The hypotheses tested in this paper, however, concern the ability to discover different phenotypes and, in particular, different local optima in the fitness landscape.
Accordingly, we want to know whether the phenotypes that are found are fitter merely because they have traits with larger magnitudes (which can receive higher fitness under the directional selection of \cref{eqn:fitness-function}) or whether they have a different pattern of traits present in the phenotype satisfying a different combination of epistatic constraints.
In particular, the GRN has $n = \lvert \symgenot \rvert$ evolvable parameters that can affect the magnitude of each trait (as well as the correlations between traits), whereas the one-to-one G--P map can only alter the elements of $\symgenot$. Accordingly, to remove any unfair advantage, our results report fitness using
\begin{equation}
  w'(\symadultphenot) = w(\theta(\symadultphenot)),
\end{equation}
where $w(.)$ is the fitness function given by \cref{eqn:fitness-function} and $\theta(x) = 1$ if $x > 0$ else $-1$.

This ignores fitness differences created by merely increasing the magnitude of a trait and instead counts the number of satisfied epistatic constraints.
It is in this sense that we refer to ``different'' phenotypes, and different local optima in the fitness landscape, i.e., disregarding differences in magnitudes that do not change the number of satisfied epistatic constraints.
This ensures that fitness improvements are due to finding different phenotypic patterns and not merely due to higher expression levels -- because it has found good phenotypes by learning what traits work well together, rather than learning to produce large traits.

\subsection*{Cross-sections through genotype space} \label{subsec:cross-sections-through-genotype-space}
To generate the cross-sections through genotype space that are used in \cref{fig:grn-cft100-cross-section-with-evolutionary-time-with-trajectories,fig:grn-cft100-cross-section-with-evolutionary-time}, we utilize the global optimum of the fitness landscape: $\symvect{T} = \langle T_1, T_2, \dotsc, T_N \rangle, T \in \{-1, 1\}^N$, where $N$ is the number of gene expression potentials.

\Cref{fig:grn-cft100-cross-section-with-evolutionary-time} utilizes a one-dimensional cross-section through a subspace of genotypes:
\begin{equation*}
\symvect{G}_i = \left\langle g_1, \dotsc, g_n \right\rangle,
\end{equation*}
where $i = \ldblbracket 0, 1, \dotsc, N \rdblbracket$. The individual values of $\symvect{G}$ are determined by
\begin{equation*}
  g_n =
  \begin{cases}
     T_n & \text{for } n \leq i \\
    -T_n & otherwise.
  \end{cases}
\end{equation*}
This cross-section maintains the property that local optima on the surfaces depicted in \cref{fig:grn-cft100-cross-section-with-evolutionary-time} are true local optima in the original high-dimensional fitness landscape also.

\Cref{fig:grn-cft100-cross-section-with-evolutionary-time-with-trajectories} utilizes a two-dimensional cross-section of similar construction, with each dimension being a particular cross-section through a subspace of genotypes:
\begin{equation*}
  \symvect{G}_\mathrm{ij} = \left\langle \langle x_1, \dotsc, x_{\frac{N}{2}}\rangle, \langle y_1, \dotsc, y_{\frac{N}{2}}\rangle \right\rangle,
\end{equation*}
where $i = j = \ldblbracket 0, 1, \dotsc, \frac{N}{2} \rdblbracket$. The first half of $\symvect{G}$ is determined by
\begin{equation*}
  x_n =
  \begin{cases}
     T_n & \text{for } n \leq i \\
    -T_n & otherwise,
  \end{cases}
\end{equation*}
and the second half by
\begin{equation*}
  y_n =
  \begin{cases}
     T_{\frac{N}{2} + n} & \text{for } n \leq j \\
    -T_{\frac{N}{2} + n} & otherwise.
  \end{cases}
\end{equation*}
Similar to the one-dimensional cross-section described earlier, the local optima on the surfaces depicted in \cref{fig:grn-cft100-cross-section-with-evolutionary-time-with-trajectories} are also true local optima in the original fitness landscape.

\section*{Acknowledgments}
We thank Rob Mills, C.\ L.\ Buckley, Adam Davies, Chris Cox and William Hurndall for discussions.
This work was supported by an EPSRC Doctoral Training Centre grant (EP/G03690X/1).
No data sets are associated with this publication.

\clearpage
\bibliography{bib-full,bibliography}

\end{document}